\pdfoutput=1
\documentclass[11pt,twoside,a4paper,cmspaper,final,collab]{cms-tdr}
\def\svnVersion{e22cb0f}\def\svnDate{2024/08/17}\def\cmsCernNoTag{CERN-EP-2023-303}\def\cmsCernDate{\today}\def\cmsMessage{Published in Computing and Software for Big Science as \href{http://dx.doi.org/10.1007/s41781-024-00124-1}{\doi{10.1007/s41781-024-00124-1}.}}
\begin{document}\cmsNoteHeader{MLG-23-001}

\newlength\cmsTabSkip\setlength{\cmsTabSkip}{1ex}

\newcommand{\TENSORFLOW} {{\textsc{TensorFlow}}\xspace}
\newcommand{\PYTORCHGEOMETRIC} {{\textsc{PyTorch-Geometric}}\xspace}
\newcommand{\PYTORCH} {{\textsc{PyTorch}}\xspace}
\newcommand{\ONNX} {{\textsc{ONNX}}\xspace}
\newcommand{\TRT} {{\textsc{TRT}}\xspace}
\newcommand{\ONNXRUNTIME} {{\textsc{ONNX Runtime}}\xspace}
\newcommand{\XGBOOST} {{\textsc{XGBoost}}\xspace}
\newcommand{\TENSORRT} {\textsc{TensorRT}\xspace}
\newcommand{\TRITON} {\textsc{Triton}\xspace}
\newcommand{\XROOTD} {\textsc{XRootD}\xspace}
\newcommand{\EXTERNALWORK} {\textsc{ExternalWork}\xspace}
\newcommand{\TRITONSERVICE} {\textsc{TritonService}\xspace}

\cmsNoteHeader{MLG-23-001}

\title{Portable acceleration of CMS computing workflows with coprocessors as a service}

\date{\today}

\abstract{Computing demands for large scientific experiments, such as the CMS experiment at the CERN LHC, will increase dramatically in the next decades. To complement the future performance increases of software running on central processing units (CPUs), explorations of coprocessor usage in data processing hold great potential and interest. Coprocessors are a class of computer processors that supplement CPUs, often improving the execution of certain functions due to architectural design choices. We explore the approach of Services for Optimized Network Inference on Coprocessors (SONIC) and study the deployment of this as-a-service approach in large-scale data processing. In the studies, we take a data processing workflow of the CMS experiment and run the main workflow on CPUs, while offloading several machine learning (ML) inference tasks onto either remote or local coprocessors, specifically graphics processing units (GPUs). With experiments performed at Google Cloud, the Purdue Tier-2 computing center, and combinations of the two, we demonstrate the acceleration of these ML algorithms individually on coprocessors and the corresponding throughput improvement for the entire workflow. This approach can be easily generalized to different types of coprocessors and deployed on local CPUs without decreasing the throughput performance. We emphasize that the SONIC approach enables high coprocessor usage and enables the portability to run workflows on different types of coprocessors.}

\hypersetup{
    pdfauthor={CMS Collaboration},
    pdftitle={Portable acceleration of CMS production workflow with coprocessors as a service},
    pdfsubject={CMS},
    pdfkeywords={CMS, offline and computing, machine learning}
}

\maketitle

\section{Introduction}
\label{sec:intro}
During the first two runs of the CERN LHC~\cite{Evans:2008zzb}, the ATLAS~\cite{ATLAS:2008xda} and CMS~\cite{CMS:2008xjf} Collaborations have analyzed trillions of high-energy proton-proton or lead-lead collisions and produced an extensive suite of physics results.
Among these are the discovery of the Higgs boson~\cite{ATLAS:2012yve,CMS:2012qbp,CMS_Higgs_2013} in the standard model (SM) and stringent constraints on various beyond the SM physics scenarios, such as supersymmetry~\cite{cms_squarks, cms_stops, cms_higgsino, cms_sleptons, atlas_squarks, atlas_stops, atlas_higgsino, atlas_sleptons} and exotic heavy-particle or dark matter candidate production~\cite{ATLAS:jetmet, CMS:jetmet, ATLAS:dijetres, ATLAS:dilepres, CMS:dijetres, CMS:highdijetres, CMS:dilepres}.
In order to measure the SM with higher precision and search for new processes with lower cross sections, the amount of data that is delivered by the LHC and processed by the experiments is expected to increase dramatically in ongoing and future physics runs~\cite{Aberle:2749422,Bruning:2015dfu}.

The high data-taking rates and increasing event complexity of the ATLAS and CMS experiments present a significant computational challenge for data processing~\cite{Zabi:2020gjd,2137107}.
A two-level trigger system is employed to run fast algorithms and reduce the data rate from 40\unit{TB/s} to about 10\unit{GB/s}~\cite{doi:10.1146/annurev-nucl-020420-093547,ATLAS:2020esi, CMS:2016ngn}.
While this is a significantly smaller rate, it is still very challenging for subsequent processing steps.
As discussed in Refs.~\cite{CMS_computing_plot,CMS:2815292,ATLAS_computing_plot},
even with optimistic expectations for computing research and development, the projected computing needs for CMS will be only narrowly satisfied.

\begin{sloppypar}
    At present, data processing is mainly carried out using central processing units (CPUs) but their expected performance increase is limited~\cite{Dennard}. Nevertheless, data processing can be supported with a variety of modern architectures, such as graphics processing units (GPUs), field-programmable gate arrays (FPGAs), application-specific integrated circuits (ASICs), or Graphcore intelligence processing units (IPUs)~\cite{Graphcore, IPU_Perf}, which can collectively be referred to as coprocessors.
    These architectures are becoming increasingly popular because of their large numbers of processing units, inherent parallelization designs, and more energy-efficient and environmentally friendly computing, especially suited for machine learning (ML) algorithm computations.
\end{sloppypar}

Within high-energy physics (HEP), deep-learning (DL) algorithms are already widely used for regression and classification tasks, and their popularity is growing rapidly~\cite{Guest:2018yhq, Albertsson:2018maf, Bourilkov:2019yoi, Larkoski:2017jix, hepmllivingreview}.
Because of this growth, inference execution for these algorithms consumes increasingly large fractions of the overall processing load.
However, the processing load can be shared and accelerated by using heterogeneous computing architectures.
Therefore, developing a framework to enable and optimize the deployment and portability of coprocessors is of considerable interest to HEP experiments~\cite{Harris:2022qtm,Savard:2023wwi}.
As a proof of concept, this paper focuses on accelerating the inference for ML algorithms in one of the CMS data processing stages, explained in Section~\ref{sec:miniaod_production}.
These algorithms collectively take about 10\% of the total processing time of that stage.
For this approach to resolve the future computing challenge, ML algorithms would need to be used for larger portions of CMS data processing.
Utilizing ML for tasks such as tracking~\cite{Farrell:2018cjr,Amrouche:2019wmx,Ju:2021ayy,DeZoort:2021rbj}, clustering~\cite{Qasim:2019otl,Kieseler:2020wcq,CMS:2022txh}, particle reconstruction~\cite{Pata:2021oez,Pata:2022wam,Mokhtar:2023fzl,DiBello:2022iwf,Pata:2023rhh}, and particle identification~\cite{Moreno:2019bmu,Qu:2019gqs,Moreno:2019neq,Bols:2020bkb,Sirunyan:2020lcu,Qu:2022mxj,CMS:2023dsu} is an active area of research.

One of the approaches for the coprocessor deployment is to equip every CPU machine with coprocessors, referred to as directly connected.
In this scenario, every CPU thread within the machine can communicate with the coprocessor.
However, since the coprocessor-to-CPU ratio must be determined before deployment, the coprocessor resources are unlikely to be optimally utilized, leading to either unused resources when undersaturating, or decreased performance when oversaturating.
In addition, it is difficult to utilize additional or more advanced coprocessor resources after deployment.

\begin{figure*}[htp]
    \centering
    \includegraphics[width=0.70\textwidth]{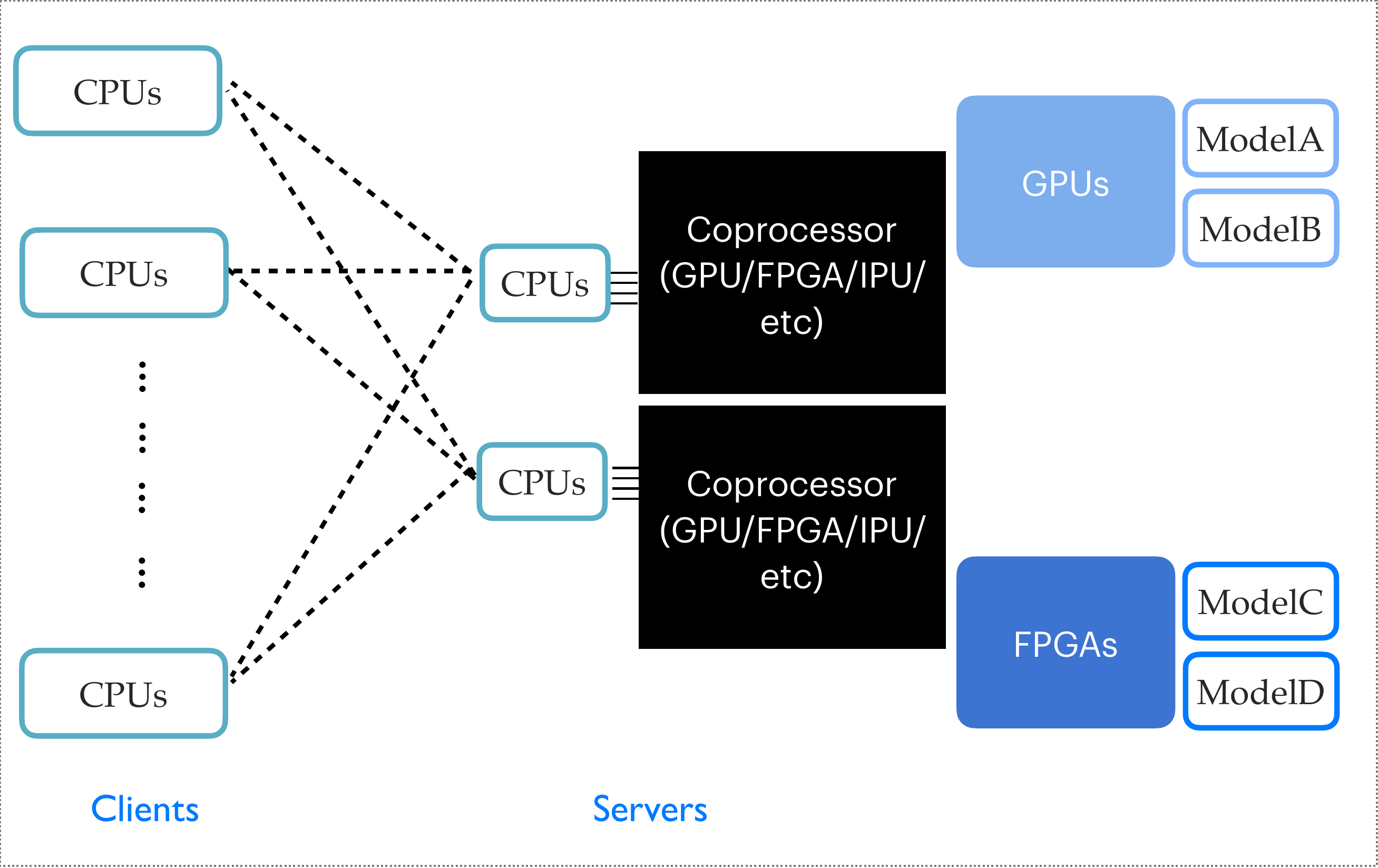}
    \caption{
        An example inference as a service setup with multiple coprocessor servers.
        Clients usually run on CPUs, shown on the left side; servers hosting different models run on coprocessors, shown on the right side.
    }
    \label{fig:illustration}
\end{figure*}

An alternative approach is inference as a service (IaaS), where coprocessor resources are separated from CPU machines.
As illustrated in Fig.~\ref{fig:illustration}, in this scheme CPU-based \emph{clients} can send the computing request with the necessary information to coprocessor-based \emph{servers} via network calls.
The servers running on coprocessor resources can perform computing tasks upon request.
This removes the restriction of a coprocessor only being used by the CPUs directly connected to it, allowing it to accept processing requests from any CPUs (local or remote), as long as network communications are available.
Certain types of coprocessors can be allocated for specific tasks, and the coprocessor-to-CPU ratio is flexible.
Resource utilization can therefore be optimized based on specific tasks, as the number of client-side jobs using a single server can be varied depending on the computational demands of a given task.
Furthermore, at the software level, since the support for coprocessors and CPU workflows is separated, it is easier to support different types of coprocessors.
This ensures algorithm portability with minimal maintenance burden.
The implementation of IaaS in experimental software frameworks can be accomplished using the Services for Optimized Network Inference on Coprocessors (SONIC) approach~\cite{Duarte:2019fta}, as described in Section~\ref{sec:sonic}.

The SONIC approach has previously been demonstrated using FPGAs~\cite{Duarte:2019fta,FaaST} and GPUs~\cite{Krupa:2020bwg,Wang:2020fjr,Cai:2023ldc,Gunny:2021gne} with a variety of ML algorithms.
These studies demonstrated that offloading ML algorithms with the SONIC approach adds little additional computational overhead from the client-server communications and other operations.
Therefore, a corresponding increase in the throughput has been observed when offloading algorithms to these faster coprocessors.

Heterogeneous computing frameworks using GPUs have appeared recently in multiple non-IaaS contexts in HEP as well.
One of the first significant real-time applications was in the ALICE high-level trigger (HLT) system~\cite{ALICE:2018phe}, where GPUs were used to accelerate track reconstruction.
Similarly, a fully GPU-based first-level trigger has been implemented in LHCb~\cite{Aaij:2019zbu}, which runs on 200 GPUs~\cite{LHCb:2023hlw}.
In CMS, a GPU-specific version of the pixel tracking domain (non-ML) algorithm called \textsc{Patatrack} was developed, along with several other reconstruction algorithms, which are now employed in the HLT for Run 3 to reduce the processing time per event~\cite{Bocci:2020pmi,Bocci:2023wmy}.
Further review of GPU usage in real-time applications for HEP can be found in Ref.~\cite{VomBruch:2020plx}.

In this paper, we take one data production workflow, called the Mini-AOD production workflow~\cite{Petrucciani:2015gjw}, as an example, and study the performance gains of applying the IaaS framework to this workflow.
The abbreviation ``AOD'' comes from a lower-level data format called Analysis Object Data (AOD) discussed in Section~\ref{sec:cmscomputing}.
In the current Mini-AOD production workflow, which is a data refinement and slimming step, about 10\% of the computing time is consumed by ML algorithm inference, which can be easily accelerated on GPUs.
We first summarize studies of the optimization and acceleration of the inference of each ML algorithm on GPUs.
Then we show that the IaaS scheme, which is implemented in the CMS software framework \CMSSW~\cite{CMS:2006myw,cmssw} via the SONIC approach, not only decreases processing time but can also be applied in large-scale production to optimize GPU utilization.
Finally, we show that the SONIC approach can be easily ported to different types of coprocessors such as IPUs, and it can also run on local CPUs without decreasing the throughput.
The study of power, power savings, and sustainability is also important, but is beyond the scope of this paper.

The paper is organized as follows.
Section~\ref{sec:cmscomputing} provides a brief overview of the CMS computing architecture and different data tiers for production.
Section~\ref{sec:sonic} discusses the SONIC approach in detail, including the technical implementation in \CMSSW, the current inference servers, and features of the approach.
Section~\ref{sec:algo} includes the data set used for the studies and the algorithms that can currently use the SONIC approach for inference.
Sections~\ref{sec:performances} and~\ref{sec:IPUs} provide detailed studies evaluating the computational performance of this approach in the Mini-AOD production workflow.
Finally, Section~\ref{sec:summary} summarizes the studies and discusses future plans.

\section{The CMS detector, software, and computing}
\label{sec:cmscomputing}
\subsection{Introduction to the CMS experiment}
\label{sec:implementation}

The LHC provides countercirculating beams of high energy protons or heavy ions, such that bunches of particles in these beams can interact with each other in the center of the CMS detector~\cite{CMS:2008xjf} nearly every 25\unit{ns}.
When particles from the countercirculating beams collide, a large variety of physical processes can occur, which lead to the creation of either fundamental or composite particles.
These particles, or their decay products, can then propagate into the CMS detector, which is designed to measure their energy and momentum.
In the context of this paper, we will refer to a readout cycle of the detector as an event.
The detector itself comprises multiple layers including silicon pixels and strips, crystal electromagnetic calorimeters, sampling hadron calorimeters, and muon spectrometers.
Each component of the detector has active elements that create electrical signals when particles interact with the detector.
Each discrete signal is called a hit.

Events of interest are selected using a two-tiered trigger system.
The first level (L1), composed of custom hardware processors, uses information from the calorimeters and muon detectors to select events at a rate of around 100\unit{kHz} within a fixed latency of 4\mus~\cite{CMS:2020cmk}.
The second level, known as the HLT, consists of a farm of processors running a version of the full event reconstruction software optimized for fast processing, and reduces the event rate to around 1\unit{kHz} before data storage~\cite{CMS:2016ngn}.
Processing in the trigger system is referred to as online computing, while subsequent processing is known as offline computing.

\subsection{The \CMSSW framework}
\label{sec:cmssw}
The \CMSSW framework is an open-source software framework that is used in triggering, data formatting and processing, simulation, and offline analysis~\cite{CMS:2006myw,cmssw}.
Components of the framework are used in combination to extract high-level physics information for each event from detector hits.
The \CMSSW framework processes each event with a sequence of algorithms, converting hits from electrical signals to position and energy measurements, linking these measurements into clusters~\cite{CMS:2020uim} and trajectories~\cite{CMS:2014pgm}, combining trajectories and clusters into single-particle representations or jets corresponding to hadronic showers~\cite{CMS:2017yfk}.
Additional algorithms in \CMSSW, \eg ML-based reconstruction algorithms, can be run to determine quantities such as an overall imbalance of the momentum in the direction perpendicular to the beam or to tag jets as containing or being produced by certain particles.

The \CMSSW framework uses Intel Threading Building Blocks~\cite{tbb} to enable task-based multithreading. As explained in Ref.~\cite{Bocci:2020olh}, this multithreading implementation allows for asynchronous nonblocking calls to external resources, such as a GPU, via \EXTERNALWORK.
This setup optimizes CPU resource utilization by minimizing downtime; the CPU is allowed to continue executing algorithms that do not require a coprocessor or depend on the results of the coprocessor-dependent algorithm while waiting for the external call to return.

\subsection{The CMS data tiers}
\label{sec:miniaod_production}

The \CMSSW framework is used both in online and offline contexts within CMS.
While the L1 trigger is hardware-based, the HLT is composed of algorithms in \CMSSW.
The framework also contains the algorithms that are used to process the raw data after they have been stored, deriving the higher-level information useful for a wide range of data analyses based on reconstructed objects.
The centralized CMS offline data processing flow involves three steps, which are performed both for raw data and simulated data sets.
In the first step, an AOD format for every event is derived.
This contains high-level information, such as reconstructed particles and jets, but the data size is large.
In the second step, a slimmed, higher-level Mini-AOD derivation is created~\cite{Petrucciani:2015gjw}.
Mini-AOD files are designed to be relatively small and accessible, serving as an intermediate step and the standard foundation for a variety of CMS physics analyses.
Finally, in the third step, a further slimmed Nano-AOD format is created that contains only very high-level physics observables~\cite{Rizzi:2019rsi}.
This format is commonly used directly for physics analyses.
Data sets are reprocessed regularly to incorporate the latest Monte Carlo event generator tunes~\cite{CMS:2019csb}, calibrations, and algorithm improvements.
Table~\ref{table:datatier} summarizes the average event size of these different data tiers with 2016--2018 (Run 2) data-taking conditions~\cite{cms_pileup}.

\begin{table}[ht!]
    \centering
    \caption{Average event size of different CMS data tiers with Run 2 data-taking conditions~\cite{Ehataht:2020ebp,Petrucciani:2015gjw,Rizzi:2019rsi}.}
    \begin{tabular}{c r r }
        Data tier & Event size [kB/event] \\
        \hline
        Raw       & 1000                  \\
        AOD       & 480                   \\
        Mini-AOD  & 35--60                \\
        Nano-AOD  & 1--2                  \\
    \end{tabular}
    \label{table:datatier}
\end{table}

In the scope of the studies in this paper, we choose Mini-AOD production as our test case.
Mini-AOD files are derived from the AOD data format, reducing the size per event by an order of magnitude.
Mini-AOD processing involves a wide variety of algorithms that propagate, filter, and reanalyze the AOD input objects.

\section{The SONIC approach}
\label{sec:sonic}
This section describes the implementation of the SONIC approach in \CMSSW and the server technology currently used, which is the NVIDIA \TRITON Inference Server (\TRITON)~\cite{triton}.
The benefits of running inference with the SONIC approach along with additional complexity and other implications are also discussed in this section.

\subsection{Implementation in \CMSSW}

\begin{figure*}[htp]
    \centering
    \includegraphics[width=0.90\textwidth]{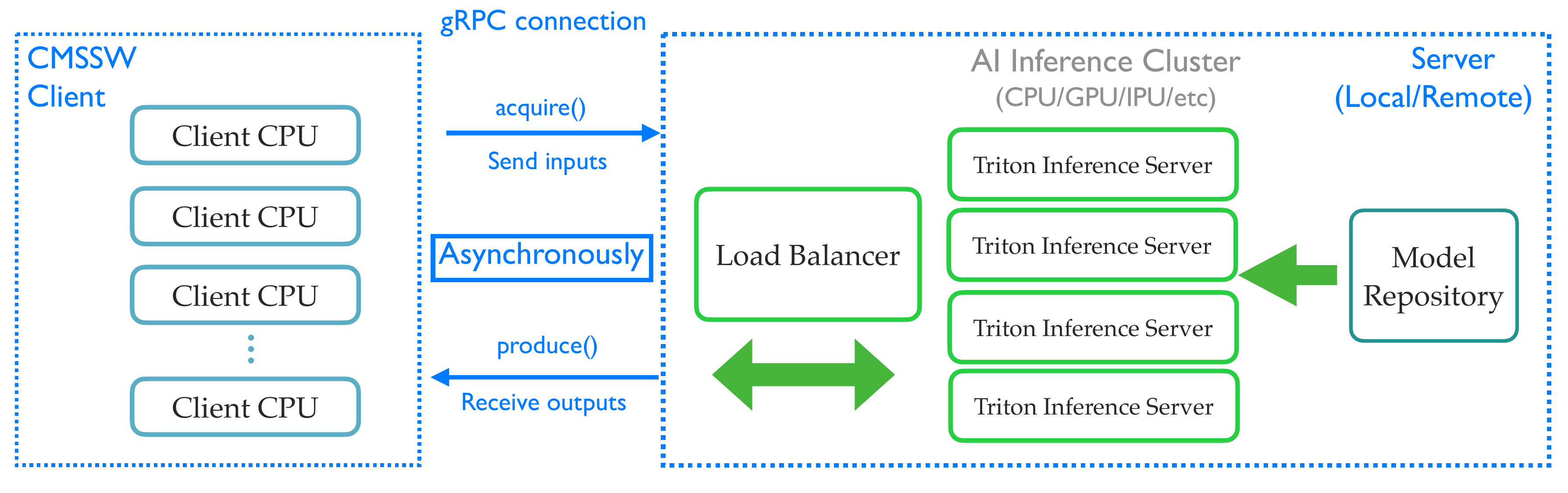}
    \caption{
        Illustration of the SONIC implementation of IaaS in \CMSSW.
        The figure also shows the possibility of an additional load-balancing layer in the SONIC scheme. For example, if multiple coprocessor-enabled machines are used to host servers, a Kubernetes engine can be set up to distribute inference calls across the machines~\cite{kubernetes}.
        Image adapted from Ref.~\cite{Wang:2020fjr}.
    }
    \label{fig:architecture}
\end{figure*}

The SONIC approach is implemented in \CMSSW through the \EXTERNALWORK framework component~\cite{Bocci:2020olh} and accesses coprocessor resources on remote servers via gRPC Remote Procedure Calls (gRPCs), which is a cross-platform open-source high-performance remote procedure call framework originally developed by Google~\cite{gRPC}.
An illustration of this procedure, where client jobs make asynchronous, nonblocking gRPC calls to remote servers, is shown in Fig.~\ref{fig:architecture}.
Multiple servers can run on multiple coprocessors, with load balancers in between. Asynchronous communication allows client CPUs to process other tasks in parallel while data are transferred between the client and the server and the inference task is processed on the server.
An important aspect of this scheme is that the client-side code does not need to be able to run any particular inference packages or frameworks; it simply has to collect the relevant input data for a trained model, communicate that information to the server in the expected format, and handle the output from the server.

The client-side framework code for the SONIC approach is split into two packages: a core package~\cite{SONIC_cms} containing class templates and other common infrastructure for the IaaS approach, and a dedicated package~\cite{SONIC_cms_triton} to interact specifically with the \TRITON server.
SONIC modules provide a similar interface to standard \CMSSW modules.
The partitioning into two packages reflects that the SONIC approach can be implemented for multiple server backends.
For example, the first implementation~\cite{SONIC_origin,Duarte:2019fta} used the Microsoft Brainwave service~\cite{Brainwave}, which provides FPGA resources.
Another potential backend service is the \TENSORFLOW as a Service framework, which can provide access to \TENSORFLOW-based ML models via the HTTP protocol~\cite{valentin_kuznetsov_2018_1308049}.
This framework was introduced along with the Machine Learning as a Service pipeline for HEP (MLaaS4HEP), which is a streamlined mechanism for training and deploying models from data in \textsc{ROOT}~data format files~\cite{kuznetsov2020mlaas4hep}.
In practice, given the generality and openness of the protocols used by \TRITON, which are themselves an extension of the KServe standard~\cite{KServe}, we expect that future development of the SONIC approach will continue to use these protocols, even if the backends or coprocessors change.
Further discussion of different backends and coprocessors can be found in Section~\ref{sec:IPUs}.

A central \TRITONSERVICE is provided to keep track of all available servers and which models each of them serves.
By default, each client-side module only needs to specify the model it needs, and the \TRITONSERVICE will automatically find a server hosting that model.
A client-side module can optionally specify a preferred server, which the \TRITONSERVICE will use if the server is found and confirmed to serve the required model.

Within \CMSSW, a mechanism has been implemented to account for the possibility that a client job cannot access a specified server for whatever reason.
In this case, a ``fallback'' server is automatically created using either local GPU resources if they are available or the CPU resources allocated to the client job in question.
The client then makes gRPC calls to that local fallback server, which introduces negligible latency.
Detailed studies related to these fallback servers are discussed later in Section~\ref{sec:fallback}.
In general, the server overhead consumes very little of the CPU resources beyond what would be used for conventional inference, such that the per-event processing time is not strongly affected by the SONIC approach relative to running without it.
Fallback servers are automatically shut down when the job finishes.

\subsection{The NVIDIA \texorpdfstring{\TRITON}{Triton} Inference Server}
\label{sec:triton}

As shown in Fig.~\ref{fig:architecture}, the server-side implementation of the SONIC approach in \CMSSW currently uses \TRITON for inference on coprocessors~\cite{triton, triton_readme}.
\TRITON is an open-source solution whose protocols are public and extensible, as noted above.
It supports inference of ML algorithms, called models, in most modern formats, including \PYTORCH~\cite{pytorch}, \TENSORRT (\TRT)~\cite{tensorRT}, \ONNXRUNTIME (\ONNX)~\cite{onnx}, \TENSORFLOW~\cite{tensorflow}, and \XGBOOST~\cite{Chen_2016}.
It also supports custom backends for alternative tasks, such as classical rule-based domain algorithms and inference on different types of coprocessors.
Several features of \TRITON are worth highlighting:

\begin{itemize}
    \item \emph{Multiple model instances}: A single server can host multiple models at the same time or even multiple instances of the same model to allow concurrent inference requests.
    \item \emph{Dynamic batching}: Usually coprocessors have many processing units and the number of operations from one inference call is not enough to fully utilize the coprocessors.
          With dynamic batching, if multiple inference calls are made within a window of time, the server can concatenate the inputs of all calls into a single batch, which improves the GPU utilization and therefore increases the overall throughput.
          In practice, the time window is typically chosen to be about the client-side CPU processing time per event, such that the inference can be processed in time without delay.
    \item \emph{Model analyzer}: Parameters like the number of model instances on a single server, the length of a batching window, and the optimal batch size are tunable and can be optimized on a case-by-case basis.
          \TRITON provides a model analyzer tool to aid in this optimization~\cite{triton_model_analyzer}.
          It can mimic the clients and send randomized inputs or some pre-saved data to the server.
          The server performance is measured, and the optimal deployment configuration can be determined by scanning the parameter space.
    \item \emph{Ragged batching}: Traditionally, inference can be performed on a batch of multiple inputs as long as each input is of the same size.
          In HEP data, the size of the input for inference can vary from one instance to another, making it harder to batch them together.
          For example, if an algorithm uses information from every particle in an event as input, it is difficult to batch inference requests from multiple events because events can have a wide range of numbers of particles.
          Ragged batching allows inference requests with different sizes to be batched together, thereby improving the performance.
          This feature is relatively new and not yet fully studied in this paper.
\end{itemize}

\TRITON servers can use one or multiple GPUs on the same machine with a built-in load balancer.
They can also run purely on CPU resources when there are no GPU resources available. For other types of coprocessors, \TRITON servers can also be used with the help of custom backends.
A server requires a trained model file and a configuration file specifying input and output variable names, shapes, types, and model versions, along with the preferred batch size and other details that can be acquired through the inference optimizations.
These model and configuration files are currently accessible through the CernVM-File System (CVMFS)~\cite{cvmfs}, and are tracked by the \CMSSW release management system.

\subsection{Advantages of the SONIC approach}
\label{sec:sonic_benefits}

Inference as a service, as implemented in the SONIC approach, provides several advantages and benefits, which are summarized here.
Many of these features arise from the differences between the IaaS approach and the more traditional approach of HEP software frameworks to use only local computing resources.
These features include:
\begin{itemize}
    \item \emph{Containerization}: The SONIC approach factorizes ML frameworks out of the client software stack, i.e., \CMSSW, reducing the workload to support a wide variety of ML models.
          With the SONIC approach, one can use any framework supported by \TRITON, including custom backends, without needing to modify the \CMSSW software stack to resolve library dependencies and ensure compatibility between all external packages.
          This allows us to pick the best inference backend for one algorithm, with less concern for the implementation details.
    \item \emph{Simplicity}: Because of the containerization discussed above, SONIC client-side code is simpler and more general than the corresponding direct inference code.
          SONIC modules need only implement the conversion of input data into the server's desired format and the reverse operation for output data.
    \item \emph{Flexibility}: In the SONIC paradigm, the connections between client CPUs and server coprocessors are not fixed.
          The servers can be physically located nearby or far from clients.
          Clients from many machines can access a single server running on either one or multiple coprocessors.
          Similarly, a single client can access multiple different servers running on multiple different machines.
    \item \emph{Efficiency}: The SONIC approach enables balanced utilization of coprocessor resources.
          By optimizing the coprocessor-to-CPU ratio for different tasks, it is easier to fully utilize coprocessor resources without oversaturating them.
    \item \emph{Portability}: Through the use of the SONIC approach, client-side code does not have to be modified to take advantage of different types of coprocessors.
          Only a consistent protocol for communicating with the inference server is required, regardless of the underlying hardware: CPU, GPU, FPGA, IPU, or any other architecture.
    \item \emph{Accessibility}: If GPUs or other coprocessors are not available locally, the only way to accelerate workflows is to access those resources remotely, as a service.
          The SONIC architecture implements this use case for CMS, allowing the use of remote coprocessors.
\end{itemize}

\subsection{Limitations and production requirements}
However, with these advantages come additional complexity and changes in resource usage.
Production jobs using the SONIC approach rely on a separate server running different software, compared to the existing scheme in which jobs only execute \CMSSW on local hardware.
This implies several additional considerations:
\begin{itemize}
    \item \emph{Server failures}: Inference servers, remote or local, may experience software or hardware failures that prevent them from running.
          These new failure modes are mostly independent from existing known sources of failures, potentially leading to an overall increase in the rate of job failures.
          However, these failures can be mitigated with server-side technology, such as the load balancer Kubernetes~\cite{kubernetes}, and client-side protocols, like the local fallback server.
    \item \emph{Network usage}: The use of remote inference servers necessarily implies an increase in network traffic, as input and output data must be communicated over the network.
          Typically, input data are much larger than output data; the total usage depends on the algorithm.
          For the Mini-AOD production workflow tests presented here, the network usage is discussed in Section~\ref{sec:scale_out}.
          High network usage can be mitigated using compression, with some tradeoff in throughput from the additional operations to compress and decompress the data.
          In the studies done here, we have not observed significant issues with network usage, and as a result, analysis of the tradeoff between compression and network usage is not included.
    \item \emph{Memory usage}: The use of remote inference servers reduces the local memory usage of production jobs, compared to the direct inference approach.
          However, the use of local fallback servers, whether to mitigate remote server failures or take advantage of the containerization and portability of the SONIC approach, implies increased memory usage.
          Running the server process locally is generally expected to use more memory than the corresponding direct inference libraries.
          Measurements of memory usage are presented in Section~\ref{sec:fallback}.
\end{itemize}
These potential drawbacks, especially uncorrelated failures and network usage, are similar to those from other distributed services used in CMS production, such as the conditions database or \XROOTD~\cite{Guida:2015gvw,Bauerdick:2012st}.
These can potentially impact the processing performance and should be studied more intensively.

Handling this additional complexity requires new components to be deployed in the CMS workflow management system.
We provide below several examples of such operational concerns.
\begin{itemize}
    \item \emph{Server discovery}: The SONIC approach allows the use of remote coprocessor resources, which requires the information of servers running on these resources, \eg, IP addresses, port numbers, served models, and number of GPUs, \etc to be available to the client jobs.
          Such information can be collected in the site configurations and provided by the job submission system or a central service.
          For the studies presented in this paper, servers are launched manually, and their addresses are written into the \CMSSW configuration files.
    \item \emph{Load balancing}: Different production jobs running on different data sets have different coprocessor resource demands.
          A load balancer, such as Kubernetes, can be set up to dynamically load and unload models on different servers and distribute client inference requests to these servers.
          The load balancing for large-scale inference requests is tested and discussed in Section~\ref{sec:scale_out}.
    \item \emph{Versioning}: CMS production requires the versions of \CMSSW, servers, ML models, and ML backends to be controlled and recorded for proper provenance tracking.
          The ML model versions are already tracked by the \CMSSW release management system; the server and backend versioning should be included in the same system for consistency.
          For the studies presented in this paper, the prebuilt \TRITON server provided by NVIDIA was used.
    \item \emph{Optimization procedure}: While the SONIC approach enables the benefits described in Section~\ref{sec:sonic_benefits}, it does not necessarily make them trivially attainable.
          Some use-case-specific optimizations are required before large-scale deployment, as explained in more detail in Section~\ref{sec:performances}.
          For example, each model should be analyzed individually to find a preferred batch size.
          Similarly, while the flexible coprocessor-to-CPU ratio permits more efficient utilization of resources, one must first determine an appropriate coprocessor-to-CPU ratio.
          This is further complicated by the fact that the ratio can depend on the hardware used, such as CPU or GPU type, and on the physics content of the data set being processed.
\end{itemize}

\section{Physics data sets, algorithms, and benchmark setup}
\label{sec:algo}
This section describes the physics data set, the ML algorithms in \CMSSW that can currently use the SONIC approach for inference, and the computing resources used in the studies.

\subsection{Data set and software versions}
In the studies, we chose to process a simulated Run 2 data set of events with one top quark and one anti-top quark (\ttbar), as it includes many types of physics objects, including leptons, heavy-flavor jets, and missing transverse momentum.
Here, a heavy-flavor jet is one that originates from a charm or bottom quark, and missing transverse momentum, or \ptvecmiss, is defined as the negative vector sum of the transverse momenta of all of the reconstructed particles in an event, and its magnitude is denoted as \ptmiss~\cite{CMS:2019ctu}.
The data set was copied to local disk to factor out the effects of remote input-output (I/O) limitations for the benchmarks. The \CMSSW version CMSSW\_12\_0\_0\_pre5 and Triton server version 21.06 were used for the studies.

\subsection{Algorithms supported by the SONIC approach}

Adapting an ML algorithm to work with the SONIC approach requires a small effort to write code to prepare the network inputs and save the network outputs within \CMSSW.
A comparison of the producers used for direct inference and for the SONIC approach can be found in Refs.~\cite{CMSSW_PN_Direct} and~\cite{CMSSW_PN_SONIC}, respectively.
Most of the pre- and post-processing steps are the same, while the operations that send inference requests to the ML backends directly or to the servers are different.

In these studies, we tested three independent and computing-intensive ML-based algorithms in the Mini-AOD workflow.
These algorithms were chosen to illustrate the performance of the SONIC approach for models with differing input and output sizes, physics applications, and backends, as detailed below.

\subsubsection{The ParticleNet algorithm}
\label{sec:PN}

Graph neural networks (GNNs) have been demonstrated to achieve state-of-the-art performance for identifying jets as arising from specific particles, a task known as jet tagging~\cite{Moreno:2019bmu,Qu:2019gqs,Moreno:2019neq}.
ParticleNet~\cite{Qu:2019gqs} is a GNN-based algorithm for jet tagging and regression that represents jets as ``particle clouds.''
This algorithm was trained in \PYTORCH~\cite{pytorch} and exported to the \ONNX format.
With the SONIC approach, it is possible to perform inference with ParticleNet in the following formats: \ONNX; \PYTORCH; and \PYTORCH with \TRT~\cite{tensorRT}, which optimizes model performance on NVIDIA devices.

There are four different trained versions of ParticleNet currently running in the Mini-AOD workflow for different purposes:
\begin{enumerate}
    \item tagging anti-\kt jets~\cite{Cacciari:2008gp}, clustered with the \FASTJET package~\cite{Cacciari:2011ma}, with a radius parameter of 0.4 (AK4 jets), abbreviated PN-AK4~\cite{CMS-DP-2023-021},
    \item tagging anti-\kt jets with a radius parameter of 0.8 (AK8 jets)~\cite{CMS-DP-2020-002},
    \item mass-decorrelated tagging for AK8 jets~\cite{CMS-DP-2020-002}, and
    \item mass regression for AK8 jets~\cite{CMS-DP-2021-017}.
\end{enumerate}
The three AK8 ParticleNet algorithms are abbreviated PN-AK8 in the following sections.
All ParticleNet variations can be hosted on \TRITON servers.

The inputs to ParticleNet are the kinematic and flavor properties of the particle constituents of each jet and the secondary vertices associated with the jet, including 20 features for one particle and 11 features for one vertex.
Up to 100 particles and 10 vertices are used in the inputs for AK8 jets; if there are more than 100 particles in a jet, the 100 particles with the highest transverse momenta are used, and if there are more than 10 vertices, then the 10 with the highest displacement from the beamline are used.
For AK4 jets, the maximum numbers of particles and vertices are 50 and 5, respectively.
For the three tagging versions of ParticleNet, the outputs of each inference are category probabilities for a variety of predefined jet categories, such as the presence of a Higgs boson or the presence of a top quark.
For the mass regression, the output is a single value: the predicted jet mass.

In Mini-AOD processing, inference is performed separately for each jet in a given event, so the number of ParticleNet inferences depends on the specific physics processes and can vary substantially from event to event.
When running the standard \CMSSW version of ParticleNet, no inference batching is performed, so each inference is truly performed separately. In this context, each inference can have a variable number of inputs with no padding involved.
When using the SONIC approach for ParticleNet, it is simplest to batch all jets in an event into a single inference request, such that input particle and secondary vertex information for each jet in an event is sent to the server in a single request.
Because ragged batching is not yet fully implemented in the inference server, the different inference requests must have the same number of inputs per batch.
Therefore, in subsequent performance studies, we adopt a maximally padded approach, where zeros are added to the vectors of particles and vertices for AK8 (AK4) jets such that they have consistent lengths of 100 (50) and 10 (5), respectively.

\begin{figure}
    \centering
    \includegraphics[width=0.45\textwidth]{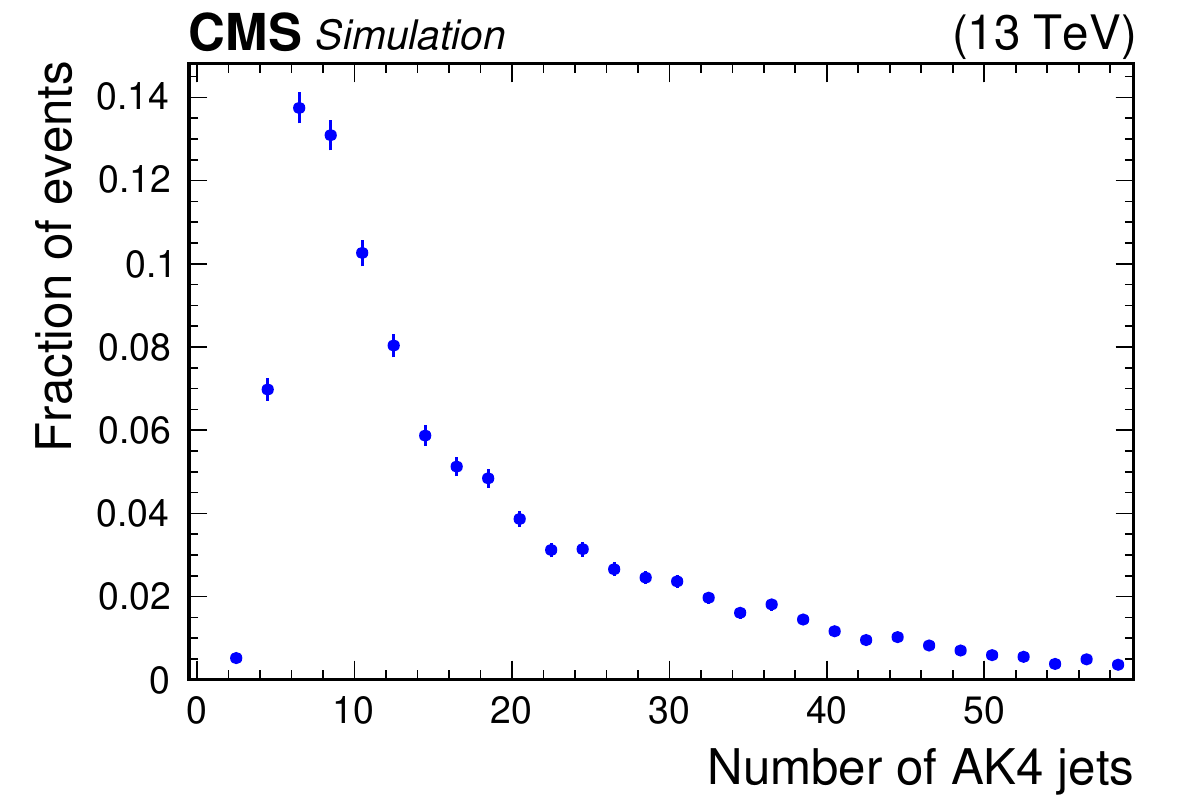}
    \includegraphics[width=0.45\textwidth]{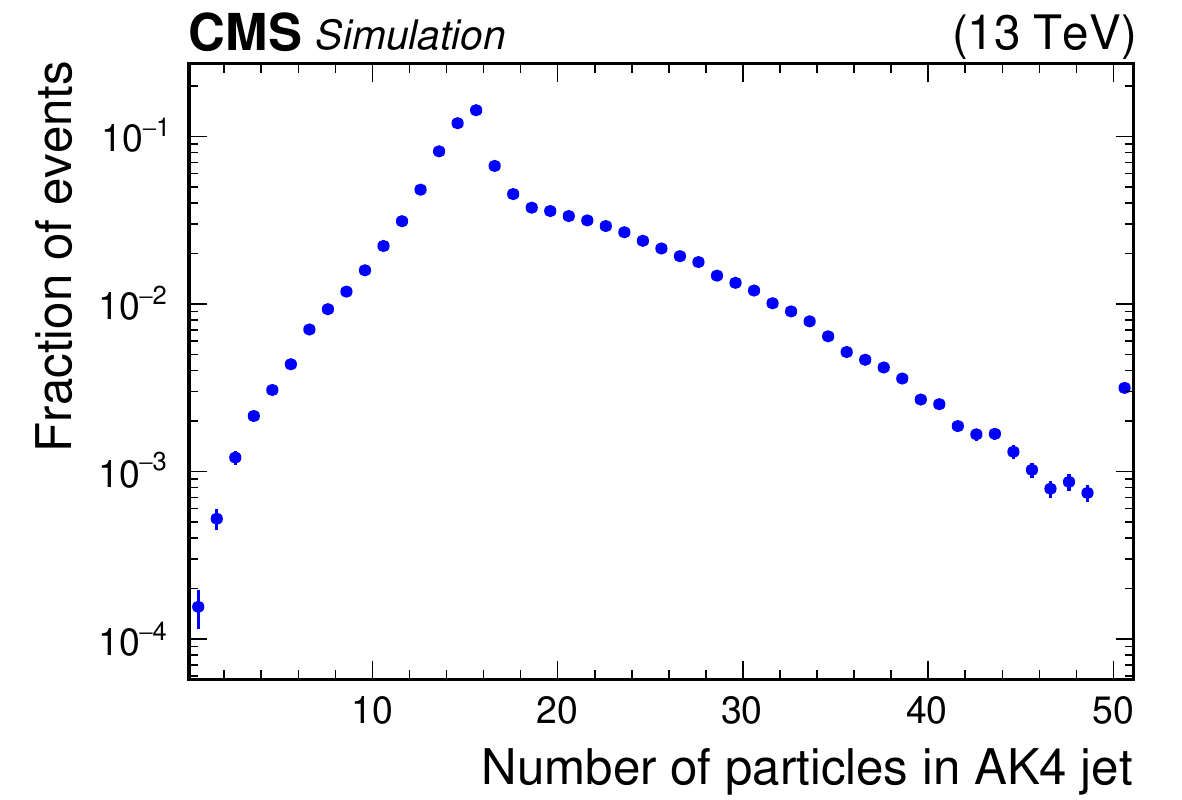}
    \includegraphics[width=0.45\textwidth]{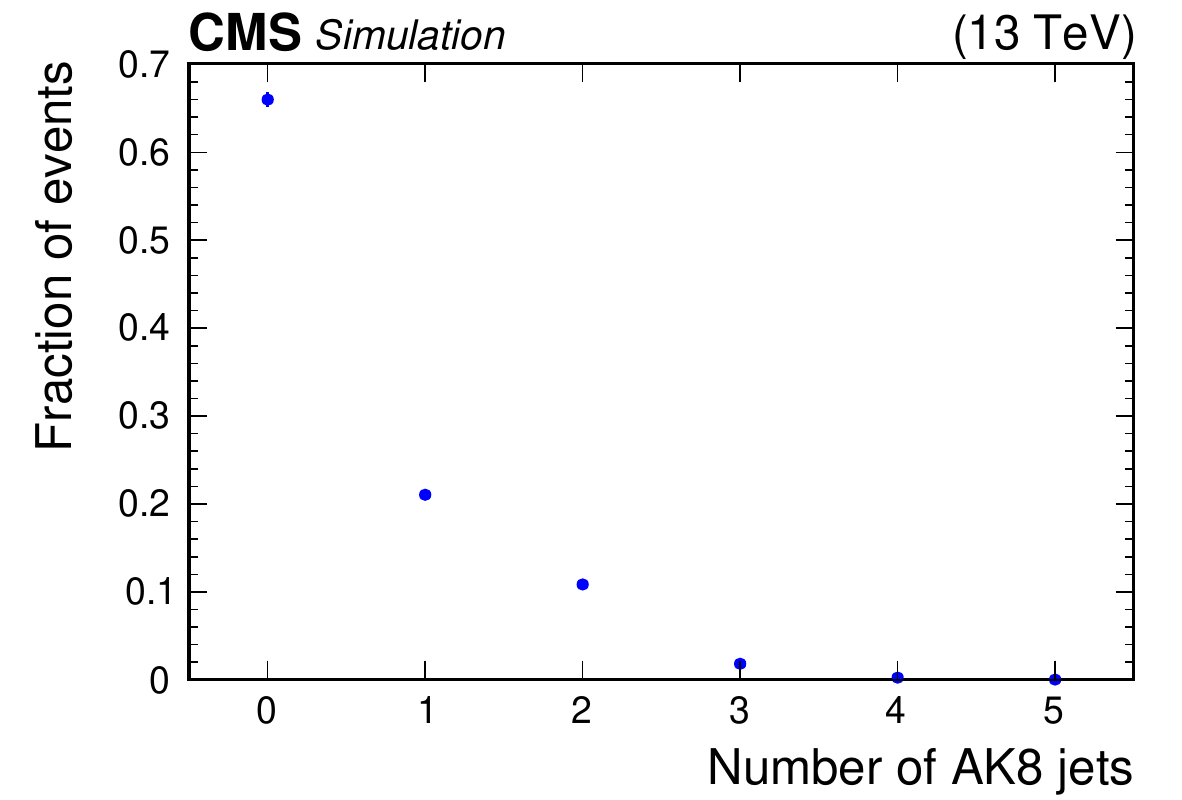}
    \includegraphics[width=0.45\textwidth]{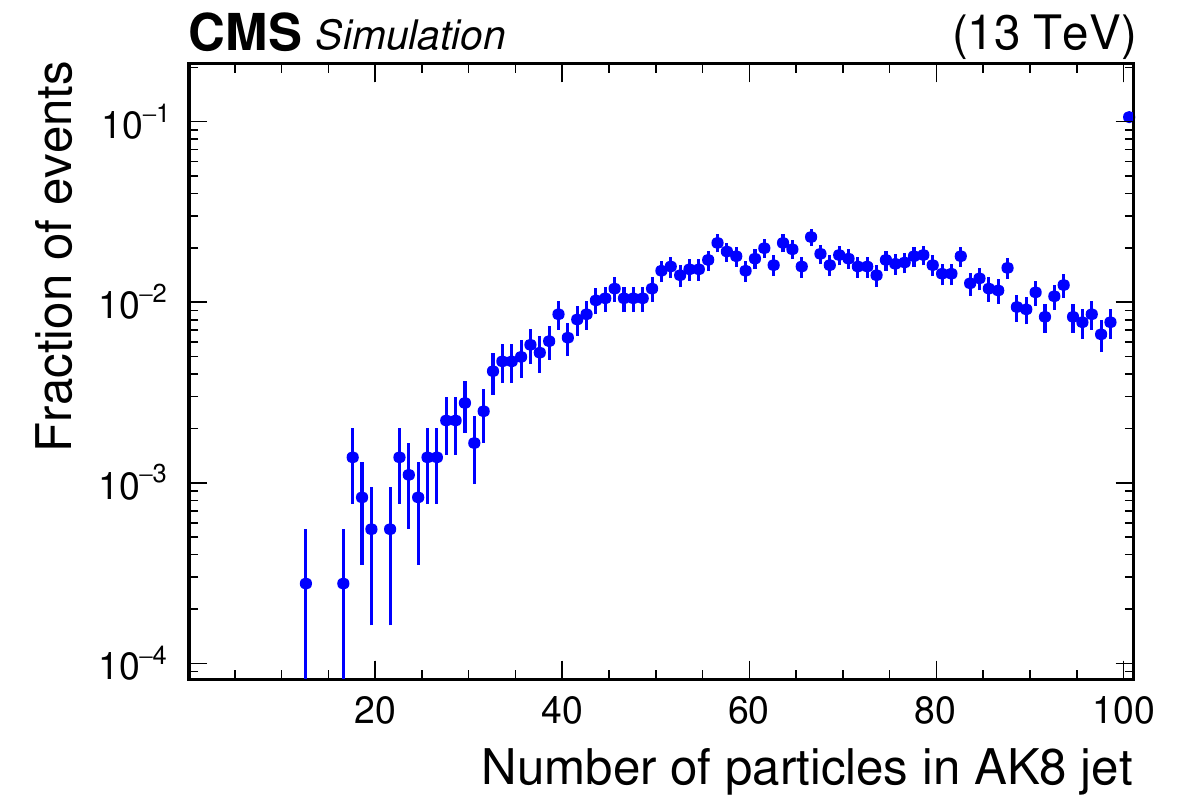}
    \caption{
        Illustration of the jet information in the Run 2 simulated \ttbar data set used in subsequent studies.
        Distributions of the number of jets per event (left) and the number of particles per jet (right) are shown for AK4 jets (upper) and AK8 jets (lower).
        For the distributions of the number of particles, the rightmost bin is an overflow bin.
    }
    \label{fig:jet_content}
\end{figure}

An illustration of the jet content of the Run 2 simulated \ttbar data set is given in Fig.~\ref{fig:jet_content}.
The distributions of the number of jets per event and particles per jet are provided for both AK4 and AK8 jets.
As shown in the figure, the number of jets can vary dramatically from event to event, indicating the importance of dynamic batching.

In the configuration where all AK4 jets are padded to 50 particles and 5 vertices and all AK8 jets are padded to 100 particles and 10 vertices, a single four-threaded job processing a Run 2 \ttbar event will generate about 140\unit{kB/s} of server input for AK4 jets and about 10\unit{kB/s} of server input for each of the AK8 jet versions of ParticleNet.
At a processing rate of about 4 events per second, with about 16 AK4 jets per event, this corresponds to about 2\unit{kB} of information per jet, which is consistent with the number of float inputs per jet.

\subsubsection{The DeepMET algorithm}
DeepMET~\cite{DeepMET} is a \TENSORFLOW-based deep neural network model that estimates the \ptvecmiss in an event.
The vector \ptvecmiss is associated with either the production of neutrinos or potential beyond the SM particles that could propagate through the detector interacting only weakly.
The inputs to DeepMET are 11 features from each particle~\cite{CMS:2017yfk} in an event, with zero-padding up to 4\,500 particle candidates for a given event.
This zero padding is used for both the standard version of DeepMET and its SONIC implementation.
In both cases, only one inference is made per event.
DeepMET outputs two values for each event: the \ptvecmiss components in the transverse plane.
Because every event necessarily has the same number of inputs, dynamic batching is automatically available via the SONIC approach, aiding inference efficiency in cases where many client jobs make concurrent requests within a certain time window (each client thread will make a request about once per second).
A single four-threaded job processing a Run 2 \ttbar event will generate about 1.3\unit{MB/s} of server input for DeepMET.

\subsubsection{The DeepTau algorithm}
The DeepTau algorithm~\cite{CMS:2022prd} is a \TENSORFLOW-based deep neural network model to identify hadronically decaying \PGt leptons from the jet collections.
The inputs to the network include low-level particle features of electrons/photons, muons, and hadrons, and high-level features such as the \PGt lepton candidate kinematic information.
The algorithm splits nearby cones into grid cells, and loops over the cells to collect the particle features.
The low-level particle features are then processed by a convolutional neural network (CNN) to extract the particle-level information.
The high-level features are processed by a fully connected neural network (FCNN) to extract the \PGt lepton candidate-level information.
The outputs of the CNN and FCNN are combined and processed by a final FCNN to produce a four-dimensional vector that represents the probability that the candidate originates from a genuine \PGt lepton, a muon, an electron, or a quark or gluon jet.
In the implementation of direct inference in \CMSSW, the network is split into three sub-models.
The first two networks process the lower-level input information separately, and the third network combines the information from the first two models with the high-level information and outputs the discriminator values.
In the SONIC implementation, we use a combined model with zero-padded inputs, so that dynamic batching can be used and GPUs can be utilized more efficiently.
A single four-threaded job processing a Run 2 \ttbar event will generate about 4.7\unit{MB/s} of server input for DeepTau, making it the most demanding algorithm explored in this study in terms of server input.

\subsection{Average processing time}

The processing time is defined as the real-world time spent between starting and finishing processing one event.
The processing time breakdown for the three algorithms highlighted above in the Run 2 \ttbar events is presented in Table~\ref{table:SONIC_Algos}, measured with 1-thread jobs using one single CPU core.
The per-event per-thread average processing time is about 1\unit{s}; PN-AK4 consumes about 4.3\% of this time and PN-AK8 about 1.1\%, while DeepMET and DeepTau take about 1.3 and 2.1\%, respectively.
Thus, for this collection of events, the algorithms supported by the SONIC approach account for about 9\% of the total processing time.
This fraction is dependent on the type of events.
For example, if there are fewer jets per event, ParticleNet will consume a smaller fraction of the total processing time.
While the fraction of the total workflow accelerated with the SONIC approach in this paper is less than 10\%, an increasing number of ML algorithms are being integrated into CMS data processing.
Because of this, it will be possible to accelerate a larger fraction of the total processing time with the SONIC approach in the future.

\begin{table*}[ht!]
    \centering
    \caption{The average time of the Mini-AOD processing (without the SONIC approach) with one thread on one single CPU core.
        The average processing times of the algorithms supported by the SONIC approach are listed in the column labeled ``Time.''
        The column labeled ``Fraction'' refers to the fraction of the full workflow's processing time that the algorithm in question consumes.
        Together, the algorithms currently supported by the SONIC approach consume about 9\% of the total processing time.
        This table also contains the expected server input for each model type created per event in Run 2 \ttbar events in the column labeled ``Input.''}
    \begin{tabular}{c r r r}
        Algorithm                     & \multicolumn{1}{c}{Time [ms]} & \multicolumn{1}{c}{Fraction [\%]} & \multicolumn{1}{c}{Input [MB]} \\
        \hline
        PN-AK4                        & 42                            & 4.3                               & 0.04                           \\
        PN-AK8                        & 11                            & 1.2                               & 0.01                           \\
        DeepMET                       & 13                            & 1.3                               & 0.33                           \\
        DeepTau                       & 21                            & 2.1                               & 1.18                           \\[\cmsTabSkip]
        PN-AK4+PN-AK8+DeepMET+DeepTau & 88                            & 8.9                               & 1.55                           \\[\cmsTabSkip]
        Full workflow                 & 990                           & 100.0                             & \NA                            \\
    \end{tabular}
    \label{table:SONIC_Algos}
\end{table*}

\subsection{Computing resources}

Fermilab, via the LHC Physics Center, provides CPU-only batch resources and a set of interactive machines with NVIDIA Tesla T4 GPUs~\cite{t4_manual}.
Through Fermilab, we were also able to steer the allocation of cloud resources (see below) using the HEPCloud~\cite{Holzman:2017jgg} framework.
These resources are physically located in Illinois.

The Google Cloud Platform (GCP) provides virtual machines (VMs) that are either CPU-only or enabled with NVIDIA Tesla T4 GPUs.
By default, the CPUs are a mix of Skylake, Broadwell, Haswell, Sandy Bridge, and Ivy Bridge architectures~\cite{intel_manual}.
In GCP, we created customized machines with specified numbers of CPU threads or different ratios of CPU threads to numbers of GPUs.
Similarly, a customized, dynamic SLURM~\cite{slurm} cluster was created that could instantiate and deplete four-thread VMs on demand for medium-scale tests that ran jobs across $\mathcal{O}(1000)$ CPUs.
The cluster's four-thread configuration was chosen so that four-threaded \CMSSW jobs would saturate the node's resources, improving the reproducibility of timing tests.
CPU-only VMs could also be instantiated through HEPCloud.
In GCP, we also maintained GPU-enabled VMs running \TRITON servers that both the SLURM and HEPCloud client nodes could access.
These resources are physically located in Iowa.

At the Purdue CMS Tier-2 computing cluster, tests were performed with reserved CPU-only and GPU-enabled machines.
The CPU-only machines are 20-core Intel E5-2660 v3 machines, and the GPU-enabled machines each have two AMD EPYC-7702 CPUs~\cite{epyc_manual}~with an NVIDIA Tesla T4 GPU.
Reserving these nodes allowed for controlled resource utilization, leading to more reproducible timing tests.
These resources are physically located in Indiana.

The diversity of resource locations used in these studies demonstrates one of this approach's key features, namely that it enables the use of nonlocal resources.
We were able to start a server in one location and have client jobs running at another site.
One such study is presented in Section~\ref{sec:different_sites}.

Within a GCP project, VMs do not have ingress bandwidth limits other than machine limits.
These are above 10\unit{GB/s}, which, based on the scans in Section~\ref{sec:algo_acceleration} and Table~\ref{table:SONIC_Algos}, is far above the amount of information that can be sent from client VMs to a server-hosting machine without saturating GPU resources.
For example, in GCP, if a single GPU is used to host all model types, then in a typical running scenario, it can service about 33 simultaneous four-threaded client jobs, or 132 client cores.
A single client core generates about 1.55\unit{MB/s} of traffic, so a total of 200\unit{MB/s} of ingress is expected per GPU at the saturation point, which is reached when the GPU is running at
the maximum throughput and cannot handle any additional incoming requests.
Similarly, the ingress bandwidth from the server into the client VMs will be very small, as inference results are no more than $\mathcal{O}(10)$ float values per algorithm.
The GCP does impose VM egress bandwidth limits, which is typically 0.25\unit{GB/s} per CPU for a VM to an internal IP address.
However, given that a single client-side core will generate about 1.55\unit{MB/s} of network traffic, this is also well within the allowed limit.
There are slightly more stringent restrictions for GCP bandwidth to external IP addresses. However, when we performed such tests as in Section~\ref{sec:different_sites}, typically only one external machine was involved, with levels of network traffic well within that allowed by the restrictions.

\section{Performance}
\label{sec:performances}
In this section, we discuss the performance of running the Mini-AOD production workflow with the SONIC approach.
We compare to direct inference, which refers to the standard approach where the inference is performed using the ML backends integrated into \CMSSW on CPUs.

As mentioned in Section~\ref{sec:triton}, we first optimize the per-algorithm workflows to find the optimal configurations for the full production workflow.
Next we check the impact of deploying servers on different sites.
Finally, we mimic the real production jobs by running scale-up tests and evaluating the performance.

\subsection{Per-algorithm inference optimization}
\label{sec:algo_acceleration}

To maximize the resource efficiency and throughput benefits of the SONIC approach, we first perform single-model characterization studies independent of \CMSSW.
For example, to maximize GPU usage without oversaturation, we need to find the optimal ratio of client-side CPUs to server-side GPUs, batch size for inference in the \TRITON server, and model configuration that will provide the highest throughput.

\begin{figure*}[ht]
    \centering
    \includegraphics[width=0.98\textwidth]{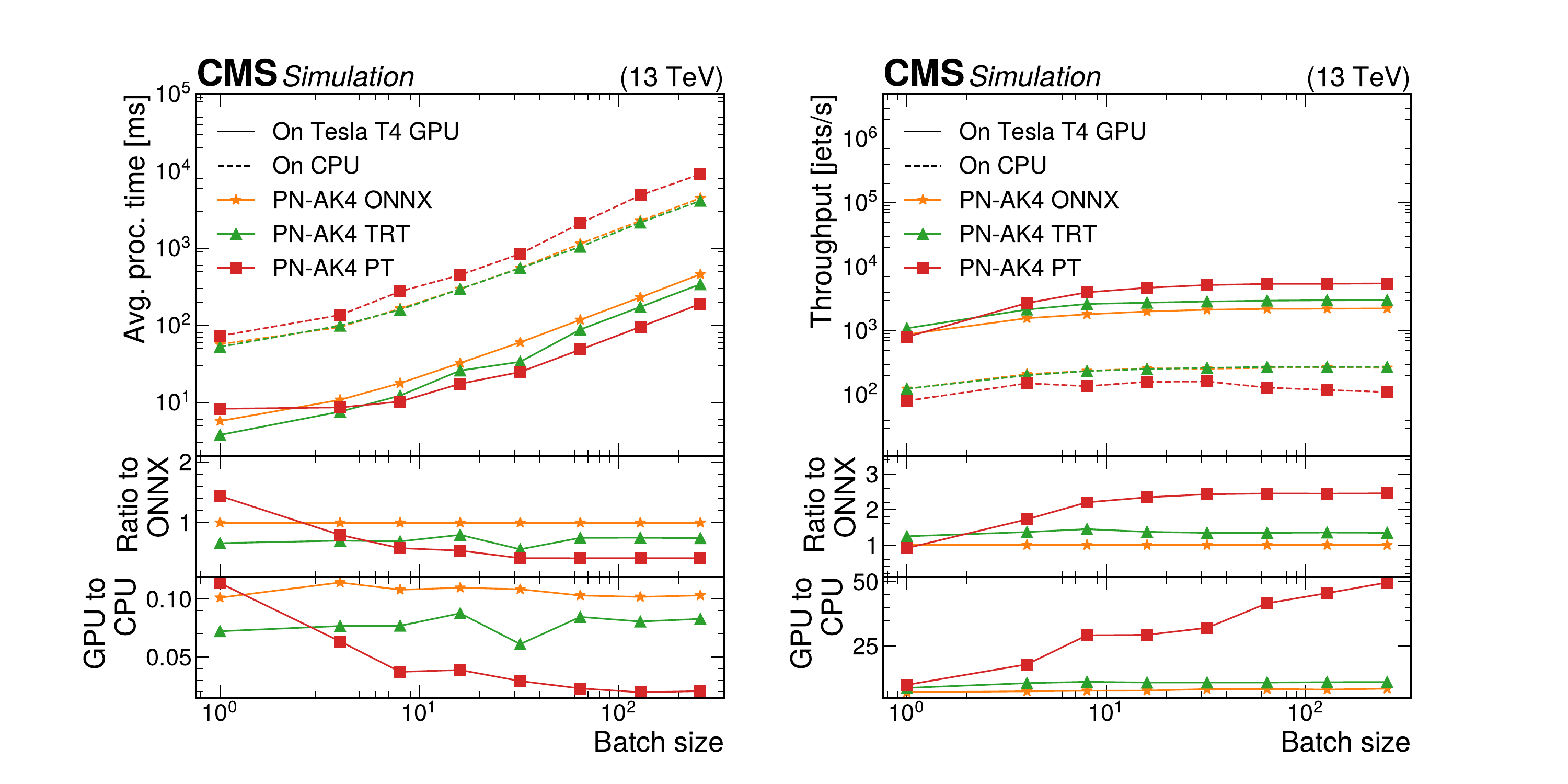}
    \caption{
        Average processing time (left) and throughput (right) of the PN-AK4 algorithm served by a \TRITON server running on one NVIDIA Tesla T4 GPU, presented as a function of the batch size.
        Values are shown for different inference backends: \ONNX (orange), \ONNX with \TRT (green), and \PYTORCH (red).
        Performance values for these backends when running on a CPU-based \TRITON server are given in dashed lines, with the same color-to-backend correspondence.
    }
    \label{fig:throughputs_pn}
\end{figure*}

The latter two optimizations can be performed with the \TRITON Model Analyzer tool~\cite{triton_model_analyzer}.
This tool feeds inputs in the correct tensor format (either randomized numbers or real data) to a loaded model hosted on a server, allowing for robust characterization of processing time per inference or exploration of the impact of batch size.
As an example, we measure the processing time and throughput of the ParticleNet algorithm for AK4 jet tagging on one NVIDIA Tesla T4 GPU, with different inference backends supported in \TRITON: \ONNX, \ONNX with \TRT, and \PYTORCH~\cite{pytorch} (labeled PT in the figures).
The results are shown in Fig.~\ref{fig:throughputs_pn}.

For smaller batch sizes, the \TRT version of the ParticleNet algorithm leads to the highest throughput in total inferences per second, while at higher batch sizes, the \PYTORCH version gives higher throughput.
In the version of ParticleNet supported by SONIC in \CMSSW, all the jets in a single event are batched together, and there are on average 16 AK4 jets per event in our \ttbar data set.
To achieve higher batch sizes in a production scenario, multiple \CMSSW clients would need to make an inference request to the same server within a relatively narrow time window.
\TRITON allows us to specify a preferred batch size, such that if many inference requests are queued within the time window, the server will perform inference with batches of approximately the specified size.
For example, the peak throughput seems to plateau around a batch size of 100 for the \PYTORCH version of ParticleNet.

\begin{figure*}[ht]
    \centering
    \includegraphics[width=0.98\textwidth]{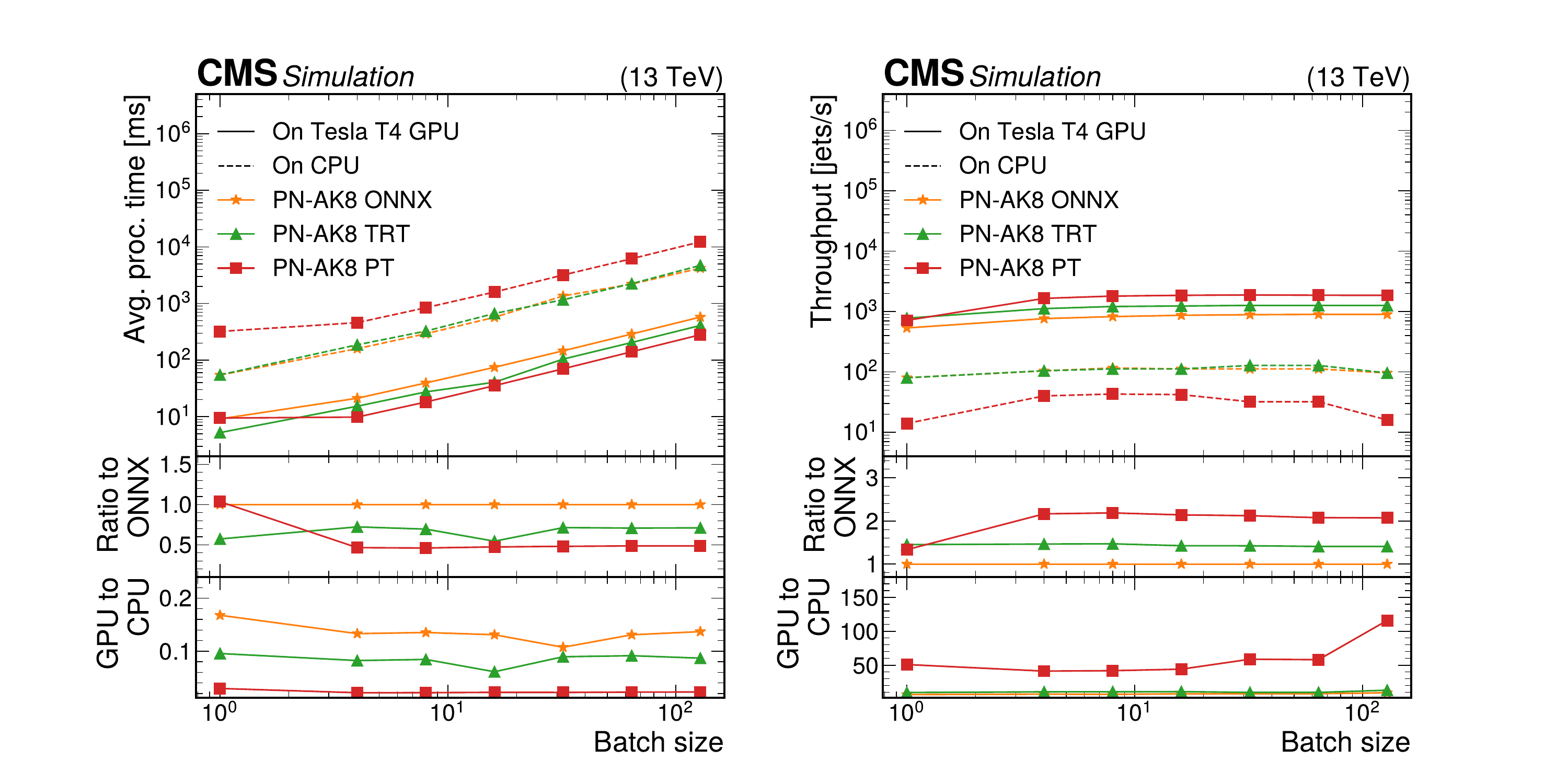}
    \caption{
        Average processing time (left) and throughput (right) of one of the AK8 ParticleNet algorithms served by a \TRITON server running on one NVIDIA Tesla T4 GPU, presented as a function of the batch size.
        Values are shown for different inference backends: \ONNX (orange), \ONNX with \TRT (green), and \PYTORCH (red).
        Performance values for these backends when running on a CPU-based \TRITON server are given in dashed lines, with the same color-to-backend correspondence.
    }
    \label{fig:throughputs_pn_ak8}
\end{figure*}

\begin{figure*}[ht]
    \centering
    \includegraphics[width=0.98\textwidth]{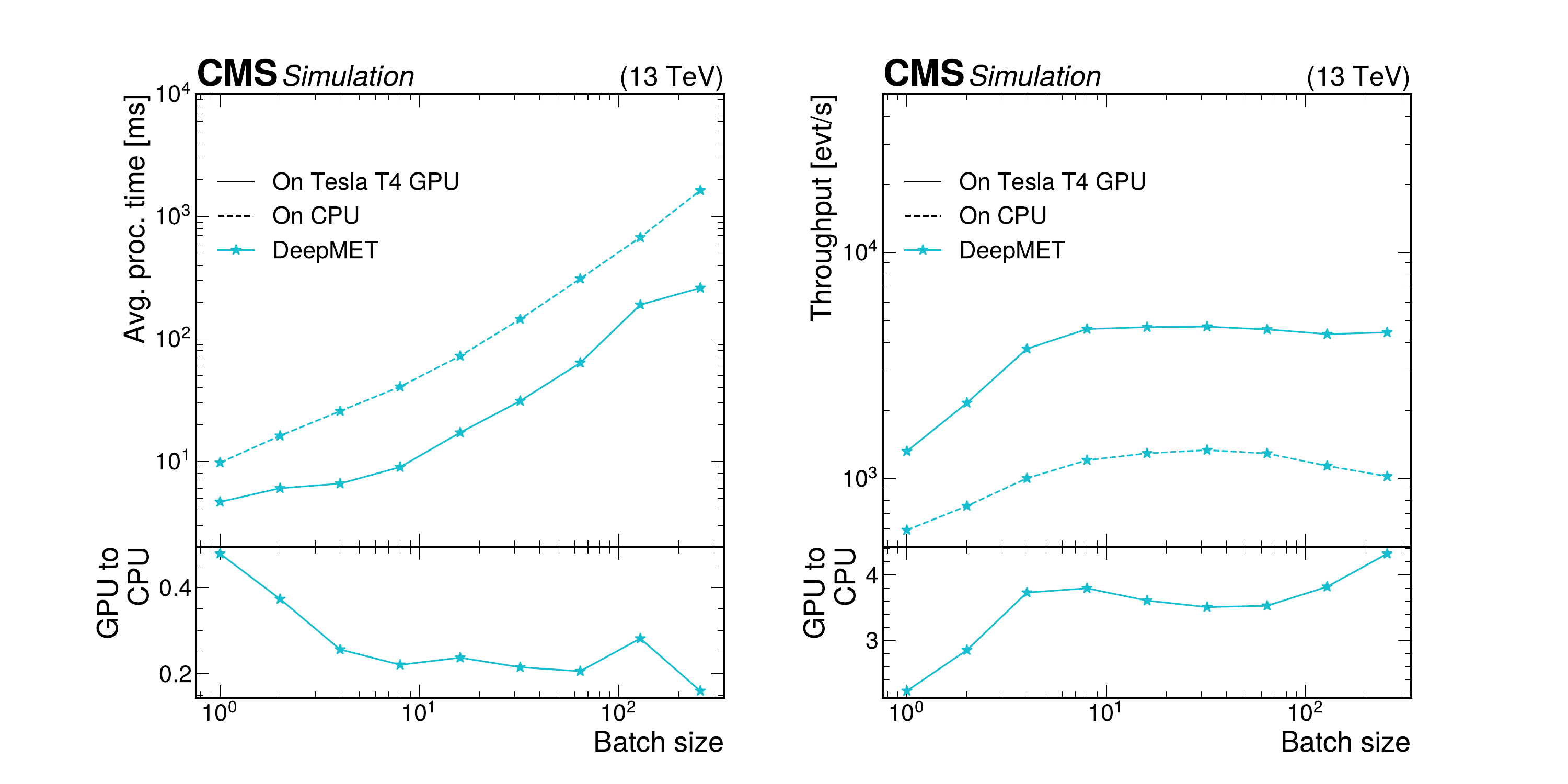}
    \caption{
        Average processing time (left) and throughput (right) of the DeepMET algorithm served by a \TRITON server running on one NVIDIA Tesla T4 GPU, presented as a function of the batch size.
        Similar performance when running on a CPU-based \TRITON server is given in dashed lines.
    }
    \label{fig:throughputs_deepmet}
\end{figure*}

\begin{figure*}[ht]
    \centering
    \includegraphics[width=0.98\textwidth]{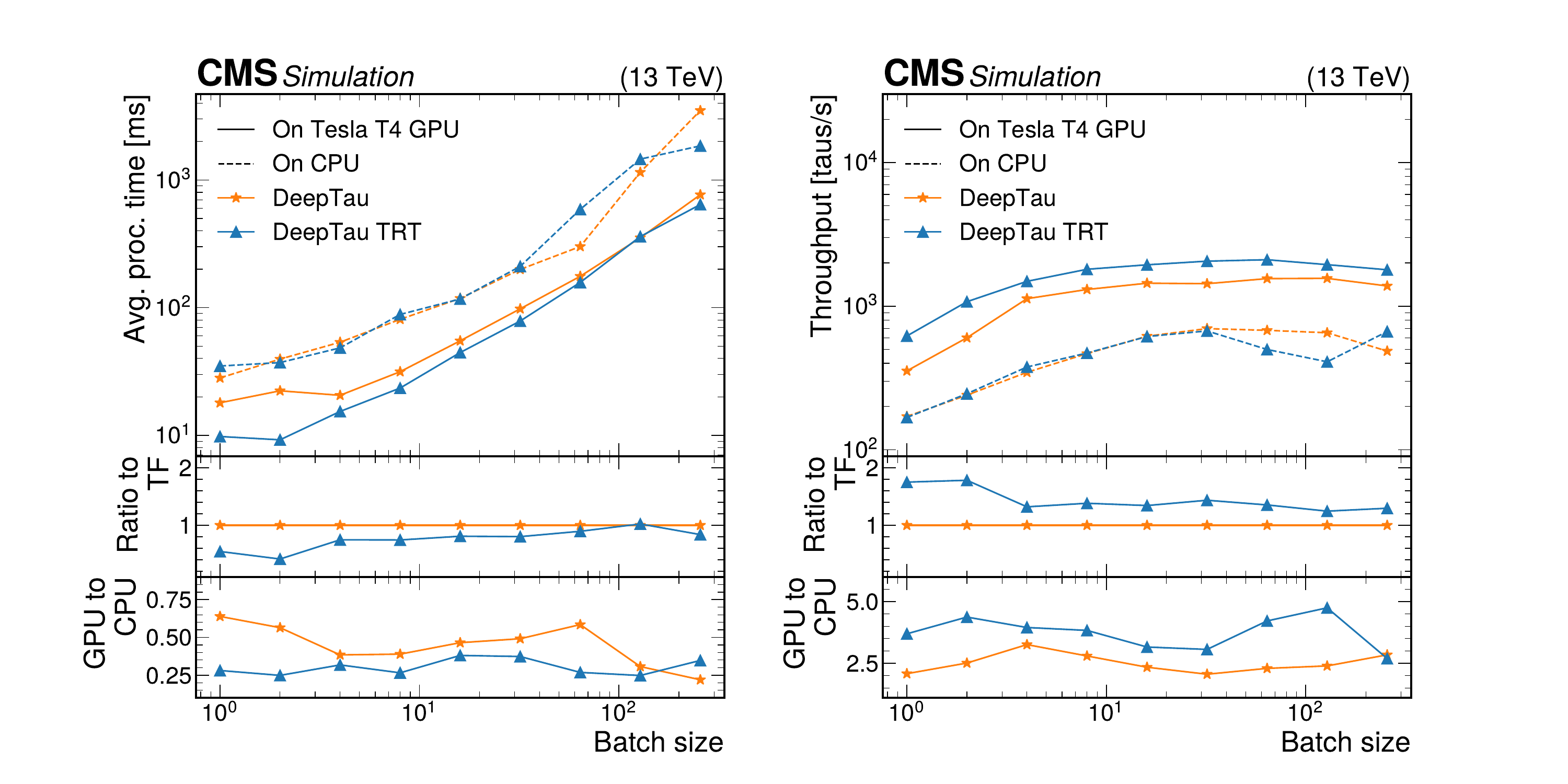}
    \caption{
        Average processing time (left) and throughput (right) of the DeepTau algorithm served by a \TRITON server running on one NVIDIA Tesla T4 GPU, presented as a function of the batch size.
        Values are shown for different inference backends: \TENSORFLOW (TF) (orange), and \TENSORFLOW with \TRT (blue).
        Performance values for these backends when running on a CPU-based \TRITON server are given in dashed lines, with the same color-to-backend correspondence.
    }
    \label{fig:throughputs_deeptau}
\end{figure*}

Similar studies can be performed on other models as well.
Figures~\ref{fig:throughputs_pn_ak8},~\ref{fig:throughputs_deepmet}, and~\ref{fig:throughputs_deeptau} show the processing time and throughput scans of PN-AK8 jet tagging, DeepMET, and DeepTau models, respectively.
Some backend tests are skipped due to the model availability.

The model analyzer can also determine approximately how many inferences per second a single GPU can perform before saturation.
For example, for the \PYTORCH version of ParticleNet for AK4 jets, a single Tesla T4 GPU can perform about 5500 inferences per second without saturating.
This corresponds to about 350 events per second, given the typical number of jets per event.
Based on this, we can estimate how many CPU clients one GPU can support in parallel.
A typical production configuration runs four-threaded Mini-AOD jobs, each of which processes about 3.9 events per second.
Therefore, a single GPU should be able to handle about 90 four-threaded jobs in parallel running asynchronously, assuming it is being used exclusively for a single server hosting the \PYTORCH PN-AK4 model.

\begin{figure}[ht]
    \centering
    \includegraphics[width=0.50\textwidth]{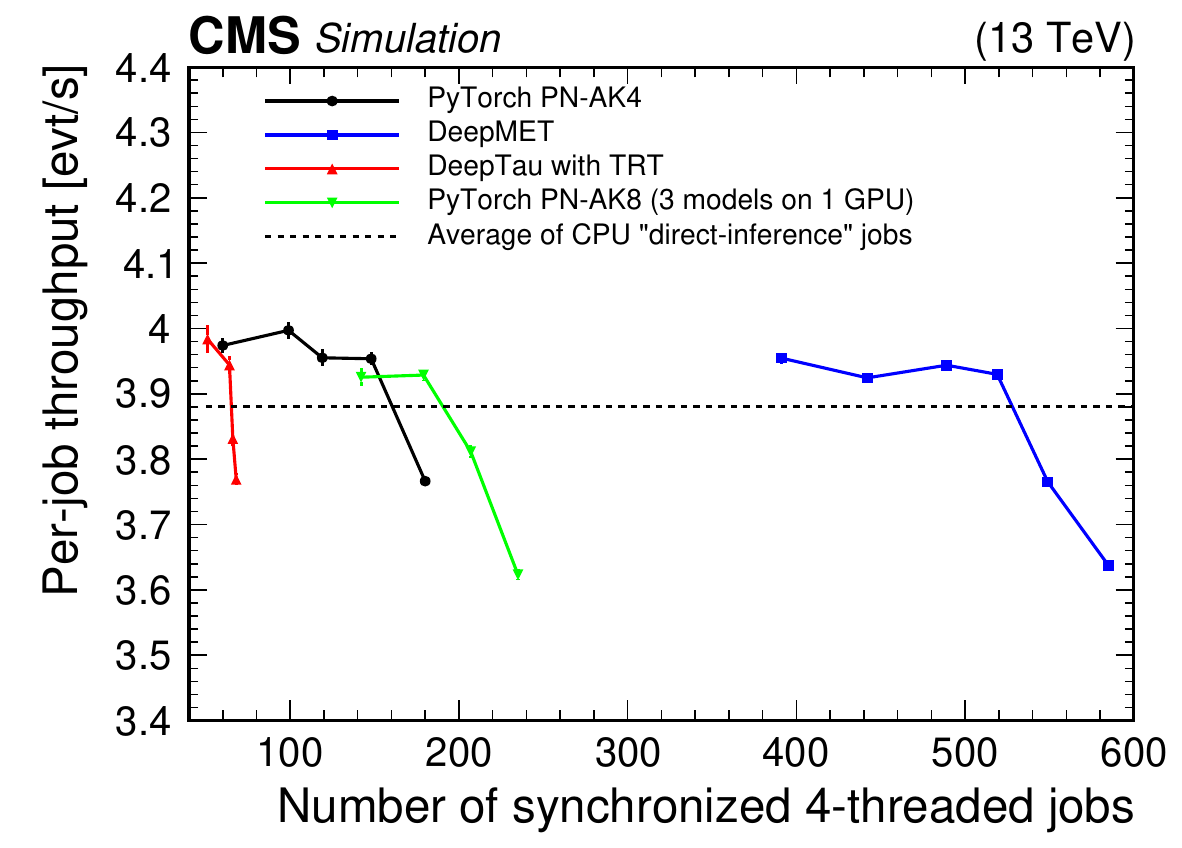}
    \caption{
        The GPU saturation scan performed in GCP, where the per-event throughput is shown as a function of the number of parallel CPU clients for the \PYTORCH version of PN-AK4 (black), DeepMET (blue), DeepTau optimized with \TRT (red), and all \PYTORCH versions of PN-AK8 on a single GPU (green).
        Each of the parallel jobs was run in a four-threaded configuration.
        The CPU tasks ran in four-threaded GCP VMs, and the \TRITON servers were hosted on separate single GPU VMs also in GCP.
        The line for direct-inference jobs represents the baseline configuration measured by running all algorithms without the use of the SONIC approach or any GPUs.
        Each solid line represents running one of the specified models on GPU via the SONIC approach.
    }
    \label{fig:throughputs_scan_pn}
\end{figure}

This expected saturation point can be tested directly by running the Mini-AOD workflow in \CMSSW and scanning the throughput as a function of the number of \CMSSW CPU clients pinging one GPU server.
Figure~\ref{fig:throughputs_scan_pn} shows such tests for the PN-AK4, PN-AK8 jet tagging, DeepMET, and DeepTau models, which were performed in GCP using a custom SLURM cluster.
For each model, a single server running on one NVIDIA T4 GPU was started on one cloud VM, and client-side jobs were executed in VMs that had 4 CPU threads.
The tests for each model class were performed separately, and as one model class was being tested, the direct-inference versions of the other models were used.

The accelerated versions of the workflow can be compared with the dashed black line, which represents the average throughput of the workflow when the direct-inference versions of all the models were used.
In this case, jobs were also started in the VMs with 4 CPU threads, but there was no communication with an external server.
As there were no shared resources between jobs, there is no expected dependence on the number of synchronized jobs.
The average processing time for this setup was determined very accurately by simply running a large number of jobs, so no associated error is shown in Fig.~\ref{fig:throughputs_scan_pn}.

When offloading PN-AK4 inference to the GPU, we expect an improvement in the overall throughput of about 4\% compared with direct inference, corresponding to the fraction of time taken by the total PN-AK4 processing shown in Table~\ref{table:SONIC_Algos}.
Such an improvement is observed before saturation, where the throughput is stable as a function of the number of simultaneous CPU clients.
The throughput decreases as the GPU starts to saturate because individual client-side jobs have to wait longer for inference requests to complete and return.
The throughput becomes lower than CPU-only inference slightly above 160 four-threaded parallel jobs, \ie, equivalent to 640 single-threaded jobs, comparable with the expected saturation point from the model analyzer.

Similar analyses were performed for the other models.
When all three variants of ParticleNet for AK8 jets are hosted on a single GPU, that GPU can serve about 190 simultaneous four-threaded client jobs.
For DeepTau and DeepMET, a single GPU hosting only one of the algorithms could serve about 64 and 520 client jobs, respectively.
The differences between these saturation points are primarily due to different model sizes and numbers of objects per event.
From these saturation values, it is possible to determine the number of GPUs needed to serve a production job that uses many client-side CPUs, and thus to determine the ratio of the number of GPUs hosting different models.
For ParticleNet and DeepTau, the saturation points will depend on the number of jets and tau leptons in the processed events, such that the ratio of GPUs hosting different models and the required ratio of GPUs to CPUs is dependent on the type of events.
If dynamic batching is enabled, the number of GPUs required for a model scales approximately linearly with the number of objects per event that require an ML inference.
For example, if there are half as many jets per event, it would require about half as many GPUs to serve ParticleNet, while the number of GPUs required for DeepMET should not change, as that model makes one inference per event.

A single GPU can host multiple models such as DeepTau and DeepMET.
In practice, it was found that loading only a single model on each GPU (split-model) led to about a 3--5\% increase in the overall performance relative to loading every model on each GPU (all-on-one).
If a 10\% improvement in throughput is observed in the split-model configuration, one would expect better than a 9.5\% improvement in the all-on-one configuration.
Because of this slight improvement, the split-model configuration was used in subsequent large-scale tests.
However, in scenarios where different physics processes are combined into a single data set, using the all-on-one configuration may be the most straightforward deployment option.
Preparatory model profiling should still be performed to ensure that enough GPUs and model instances will be available, but the difference in performance between model partitioning schemes is small.

\subsection{Cross-site tests}
\label{sec:different_sites}

\begin{figure}[ht]
    \centering
    \includegraphics[width=0.50\textwidth]{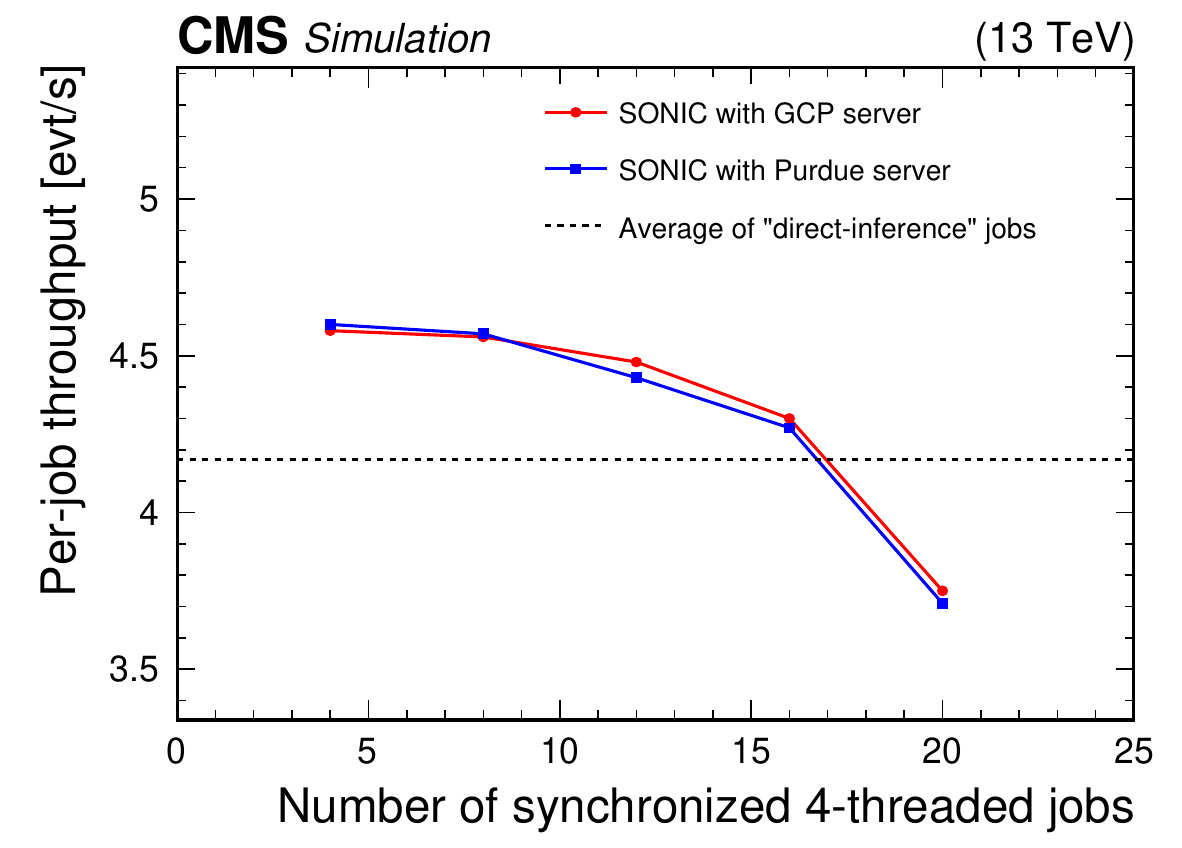}
    \caption{
        Production tests across different sites.
        The CPU tasks always run at Purdue, while the servers with GPU inference tasks for all the models run at Purdue (blue) and at GCP in Iowa (red).
        The throughput values are higher than those shown in Fig.~\ref{fig:throughputs_scan_pn} because the CPUs at Purdue are more powerful than those comprising the GCP VMs.
    }
    \label{fig:crosssites}
\end{figure}

A potential bottleneck in the SONIC paradigm is the increased inference latency caused by the physical distance between client and server and other network traffic.
While a previous study observed that the average processing time difference between remote and on-premises servers is negligible~\cite{Krupa:2020bwg}, we tested this observation explicitly with the Mini-AOD workflow, with results presented in Fig.~\ref{fig:crosssites}.
In this test, client-side jobs were executed at Purdue's Tier-2 computing cluster in Indiana.
All the models are loaded into one server running on a single GPU for simplicity.
The blue points and lines show the throughput improvement in the Mini-AOD workflow when the client jobs communicate with \TRITON servers hosting all the models at the same time, running on a single GPU also physically located at Purdue.
The improvement is shown as a function of the number of simultaneous four-threaded client-side jobs running at once.
The single GPU server begins to saturate when about 10 client-side jobs are sending requests at once and the Mini-AOD workflow running with the SONIC approach becomes slower than the CPU-only workflow if more than about 17 four-threaded client-side jobs are running at once.
The direct-inference line in Fig.~\ref{fig:crosssites} was made in the same way discussed in Section~\ref{sec:algo_acceleration} for Fig.~\ref{fig:throughputs_scan_pn}, though the CPU-only jobs were run at Purdue rather than GCP in Iowa.

These cross-site tests were performed before model configuration optimization as discussed in the preceding section.
The exact model configuration is less important for this test, as long as the near and far servers have the same type of GPUs and host the same models.
Here, the non-\TRT version of DeepTau was used, which saturated close to 40 synchronized four-threaded jobs.
When all the models are loaded on a single GPU server, the approximate saturation point can be found with reciprocal addition.
With the saturation points of 90 (AK4 jet ParticleNet), 190 (AK8 jet ParticleNet, 3 models), 40 (DeepTau), and 520 (DeepMET, 2 models), the expected saturation point is around 18 synchronized four-threaded jobs, when running all the models together on one GPU.
As noted in the previous section, this estimate is often about 5\% high.
The saturation point seen in Fig.~\ref{fig:crosssites} is lower than expected from the above analysis due to the different configurations of the CPU machines in each scenario.
Twenty-threaded, hyperthreading-disabled Intel Xeon Processor E5-2660 v3 units were used at Purdue, which are more powerful processors than those used in the GCP-based tests.
The faster the client-side resources process events, the lower the saturation point is for a given type of GPU.

The blue points and line in Fig.~\ref{fig:crosssites} show the throughput improvement when the client-side jobs are once again run at Purdue, while the \TRITON server hosting all the models on a single GPU is run on GCP resources, which were physically located in Iowa for this test.
Both the observed throughput increase and observed GPU saturation point are about the same for both server locations, so we conclude that the client-to-server distance has little impact on performance up to a few hundred kilometers.
In the future, it will be important to monitor the impact of distance, especially beyond a few hundred kilometers.

\subsection{Large-scale tests}
\label{sec:scale_out}

Finally, we perform large-scale tests to emulate realistic Mini-AOD production scenarios.
These tests were performed exclusively using GCP resources.
Here, 24 NVIDIA Tesla T4 GPUs were used to host the \PYTORCH version of ParticleNet for AK4 jets, 20 GPUs were used to host the \PYTORCH version of all three ParticleNet models for AK8 jets, 48 GPUs were used to host DeepTau, and 10 GPUs were used to host DeepMET.
The ratios of the number of GPUs hosting each model do not exactly match those expected from Section~\ref{sec:algo_acceleration}; this was done for multiple reasons.
First, a larger number of GPUs than strictly necessary were used in the tests to ensure that the performance would meet the expectations described in Section~\ref{sec:algo_acceleration}, and thereby avoid the cost of repeating the test multiple times.
For instance, based on a single-GPU saturation point of 150 four-threaded jobs, one would expect that only about 17 GPUs would be needed to serve the AK4 jet ParticleNet model.
It is worth noting that while the demonstration illustrated here uses a CPU-to-GPU ratio of 9820:102 (about 1 GPU per 96 CPU cores), achieving a higher ratio is likely possible as we could have safely decreased the number of GPUs used.
Second, it was more practical to instantiate VMs with factors of 4 GPUs.
This had the added benefit of maximizing the allowed I/O bandwidth for VMs, which in GCP is restricted for VMs with fewer CPU cores.
A maximal CPU-to-GPU ratio for VMs in GCP is achieved for machines with 4 GPUs.
While I/O bandwidth was not expected to create problems for this large test, a conservative approach was taken to mitigate the risk of needing repeated trials.
Thus, the GPU allocation approach was to find the minimum number of GPUs expected based on single saturation scans (17 for the AK4 jet ParticleNet model, 14 for the AK8 jet ParticleNet models, 42 for DeepTau, and 5 for DeepMET), find the next largest number divisible by 4, then add one extra server corresponding to 4 additional GPUs.
For DeepMET, two 4-GPU VMs were used along with one 2-GPU machine, as DeepMET is a relatively lightweight algorithm, and the approach used for the other algorithms would have more than doubled the number of GPUs allocated for the algorithm.

A separate Kubernetes load-balancer was used for each model type to distribute inference requests evenly among server-hosting, GPU-enabled VMs.
Thus, client VMs used separate IP addresses for each model type, and each inference for a particular model type was passed through a single load-balancing machine, allowing for network monitoring for each model separately.

Client-side jobs also ran on CPU-only GCP resources, using HEPCloud to dynamically allocate preemptible resources and assign jobs to the client-side VMs.
Each client job was run in a four-threaded configuration, with input data files stored locally, and each VM created in this HEPCloud setup had 32 cores and 160\unit{GB} of memory, meaning up to 8 simultaneous jobs could run in a single VM.

The largest test had 2500 simultaneous client-side jobs, which amounts to 10\,000 CPU cores.
Because these jobs were run on preemptible resources, Google reserves the right to reallocate any VM to higher-priority requests from other GCP users.
Of the 2500 jobs, 2455 jobs completed successfully without preemption, so in total 9820 client-side CPUs were used.

\begin{figure}[ht]
    \centering
    \includegraphics[width=0.50\textwidth]{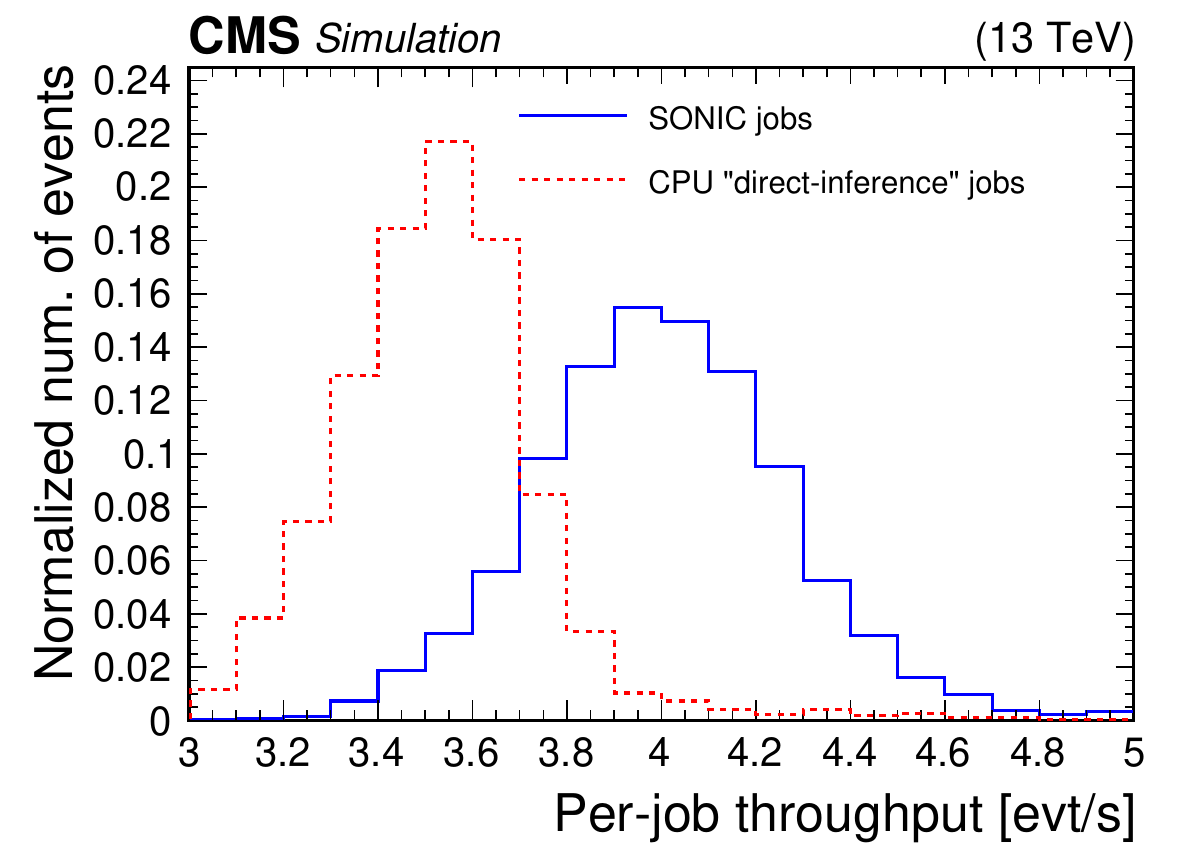}
    \caption{
        Scale-out test results on Google Cloud.
        The average throughput of the workflow with the SONIC approach is 4.0\unit{events/s} (solid blue), while the average throughput of the direct-inference workflow is 3.5\unit{events/s} (dashed red).
    }
    \label{fig:scaleout}
\end{figure}

The results of this large-scale test are shown in Fig.~\ref{fig:scaleout}.
The jobs running with the SONIC approach achieved an average throughput of 4.0\unit{events/s}, while CPU-only benchmarking jobs had a throughput of 3.5\unit{events/s}.
This 13\% increase in throughput matches the expectation of completely removing the ParticleNet, DeepTau, and DeepMET inference from the total Mini-AOD per-event processing time within the uncertainty, which is typically around 3\% from small-scale tests.
The throughput values for this test are slightly different from those shown in Figs.~\ref{fig:throughputs_scan_pn}~and~\ref{fig:crosssites}.
As noted previously, this is due to the difference in the CPUs used in the tests.

\begin{sloppypar}
    As mentioned before, server-side VMs were optimized to allow maximal input and output bandwidth.
    No bottlenecks due to bandwidth were observed in this scale-out test.
    We noted that the maximum data input rate received by one of the Kubernetes load-balancers was 11.5\unit{GB/s}, which was for DeepTau.
    The next-highest data input rate was 3.3\unit{GB/s} for DeepMET, and less than 500\unit{MB/s} of input was needed for all ParticleNet models combined.
    These values are consistent with expectations from Table~\ref{table:SONIC_Algos}, as there were roughly 10\,000 CPU cores running simultaneously, each processing about 1 event per second.
    The output rate was significantly smaller for each model, as most return only one or a few floating-point values as the inference result.
    In the future, more algorithms will be able to be run on coprocessors.
    These could have larger I/O sizes than the algorithms considered here, so continuing to monitor the network usage will be important.
\end{sloppypar}

\section{Portability}
\label{sec:IPUs}
While the inference servers discussed so far have exclusively utilized GPU resources, servers are easily portable and can run on other processing platforms.
Previous uses of the SONIC approach with FPGAs are reported in detail in Refs.~\cite{Duarte:2019fta,FaaST}, where the portability was demonstrated by changing the coprocessor technology and server backend without modifying the client-side software.
In the new studies reported here, we have additionally run the Mini-AOD workflow with servers using both CPUs and Graphcore IPUs~\cite{Graphcore, IPU_Perf}.

\subsection{The CPU fallback server}
\label{sec:fallback}

When running with remote servers, one potentially common and important failure mode is communication errors between clients and servers.
To support automatic local CPU inference as a backup option when communication failures occur with remote servers, the SONIC implementation includes a service that can launch a \TRITON server using local CPU resources for any SONIC approach-compatible models, referred to as a fallback server.
Fallback servers can also be used for inference when third-party ML frameworks are not supported for direct inference in \CMSSW.

Ideally, the use of fallback servers should have minimal impact on per-event throughput relative to running direct-inference jobs without the SONIC approach.
This is contingent on two factors.
First, the latency introduced by sending data to or from local servers must be negligible.
Second, servers should introduce minimal overhead to maximize the CPU resources used to perform inference.
The first concern can be resolved using the shared-memory option, which skips the gRPC communication and directly passes the data in certain memory chunks between the server and client.
The gRPC overhead in most cases is found to be negligible.
Regarding the second concern, local servers are running on the same CPUs as the other modules in the workflow, so scheduling efforts must be made to avoid CPU thread contention between the two.
This implies that the synchronous server mode is preferred for local CPU fallback servers.
Additionally, inference tasks should not create extra threads to avoid contention.
In our experiments, we found that using more inference threads in the server than the number of threads allocated for the \CMSSW job will slow processing down dramatically.

\begin{figure}
    \centering
    \includegraphics[width=0.50\textwidth]{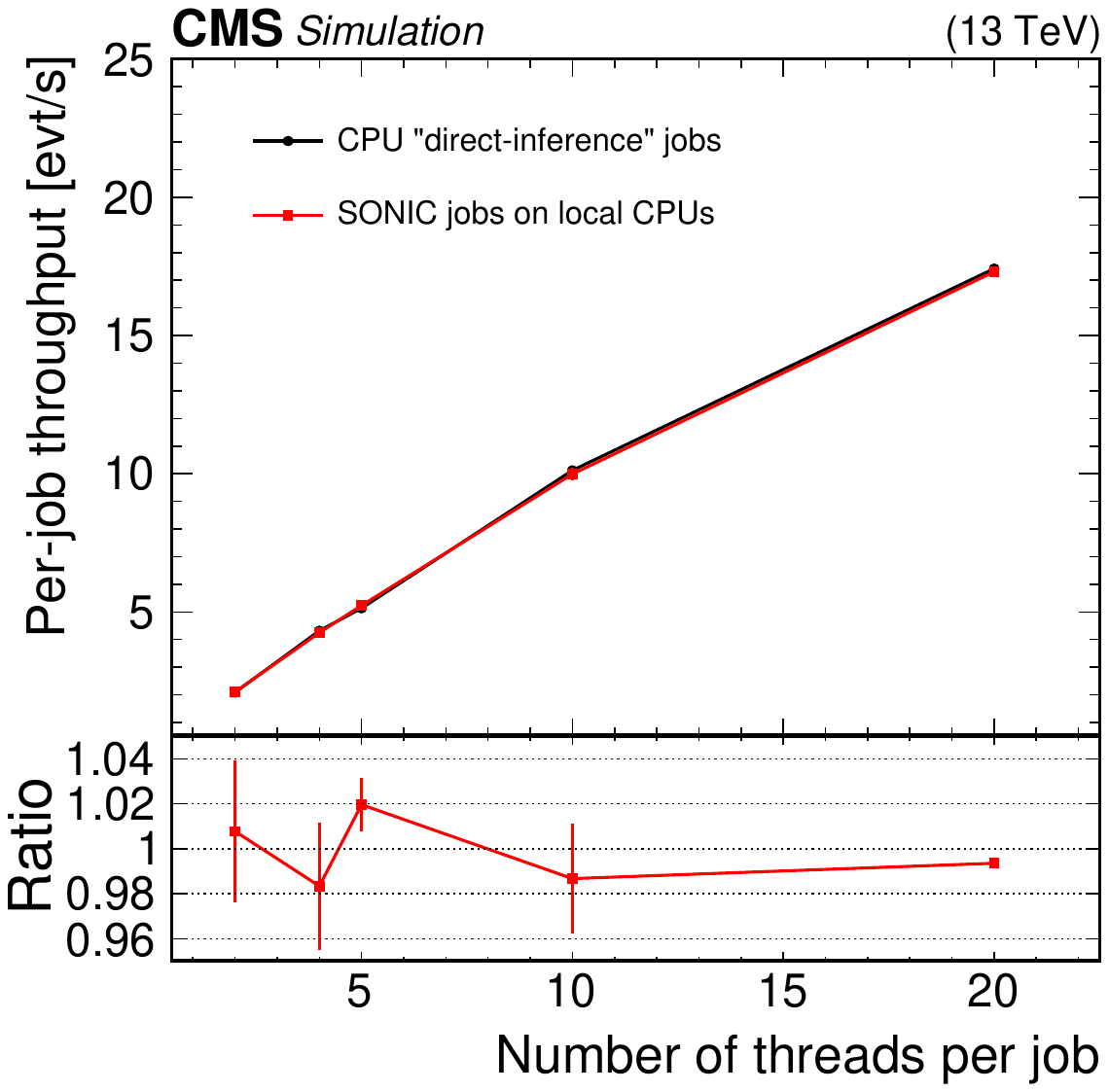}
    \caption{
        Throughput (upper) and throughput ratio (lower) between the SONIC approach and direct inference in the local CPU tests at the Purdue Tier-2 cluster.
        To ensure the CPUs are always saturated, the number of threads per job multiplied by the number of jobs is set to 20.
    }
    \label{fig:throughput_cpu}
\end{figure}

Having thus explored potential configurations, for the local CPU inference, we run our servers in synchronous mode, with the number of model instances set to the number of threads per job, and the number of inference threads always set to one.
This configuration mimics direct inference and avoids thread over-subscription as much as possible.
We compare the throughput between direct inference and the SONIC approach with local CPU fallback servers using this configuration.
Tests were performed using resources at the Purdue Tier-2 cluster with the CPU-only nodes.
There are $n_{\text{CPU}}=20$ Intel E5-2660 CPU cores on one node, and hyperthreading is disabled to ensure more stable results.
For all tests, we always saturate the CPU nodes by requiring the product of the number of jobs ($n_\text{j}$) and number of threads per job ($n_\text{T}$) to be equal to the number of CPU cores: $n_\text{j}n_\text{T} = n_{\text{CPU}} = 20$.
For example, to test a configuration where a single job occupies four threads ($n_\text{T} = 4$), we would run five synchronized jobs ($n_\text{j} = 5$), while a two-threaded configuration would require 10 synchronized jobs.
We scan the throughput as a function of the number of threads, as shown in Fig.~\ref{fig:throughput_cpu}.
The throughput of running with local CPU fallback servers is similar to direct inference.
The higher throughput in some cases is a result of optimizations in more recent versions of \ONNXRUNTIME installed on the server, which can be controlled in real production jobs.

Memory usage was also monitored with the \textsc{top} command during the studies and provided in Table~\ref{table:fallback_memory}.
As the number of threads per job increases, the number of model instances increases as well, leading to higher memory usage in the SONIC approach compared with direct inference.
On the other hand, if the number of model instances is fixed to one, the server memory usage is always around 300\unit{MB}, so the total memory usage is similar to direct inference, but the throughput decreases by 5--10\%, depending on the tasks.
Further simultaneous optimization of the throughput and memory usage will be explored in the future.

\begin{table*}
    \centering
    \caption{
        Memory usage with direct inference and the SONIC approach in the local CPU tests at the Purdue Tier-2 cluster.
        The last column is calculated as the sum of client and server memory usage divided by the direct inference memory usage.
        To ensure the CPUs are always saturated, the number of threads $n_\text{T}$ per job multiplied by the number of jobs is set to 20.
    }
    \label{table:fallback_memory}
    \begin{tabular}{c c c c c}
        $n_\text{T}$ & \CMSSW with           & \CMSSW with SONIC  & SONIC app.  & \multirow{2}{*}{SONIC/direct} \\
        per job      & direct inference [MB] & app. (client) [MB] & server [MB] &                               \\
        \hline
        1            & 1850                  & 1700               & 200         & 103\%                         \\
        2            & 2000                  & 1800               & 400         & 110\%                         \\
        5            & 2200                  & 1950               & 800         & 125\%                         \\
        10           & 2500                  & 2200               & 1200        & 136\%                         \\
        20           & 2900                  & 2500               & 2000        & 155\%                         \\
    \end{tabular}
\end{table*}

\subsection{Studies with Graphcore IPUs}

As discussed in Sections~\ref{sec:triton} and~\ref{sec:sonic_benefits}, NVIDIA \TRITON inference servers support custom backends to run with different coprocessors and different (\eg, ML) backends.
Since the SONIC approach's client code only depends on the \TRITON protocols, algorithms implemented in this way can easily be ported to different types of coprocessors.
One of the Mini-AOD production tests was run together with the Graphcore IPU team, where a custom backend was prepared by the developer team to support running ML inference with IPUs.
The supported ML frameworks on IPUs when the tests were performed included \TENSORFLOW, \ONNX, and \PYTORCH. \PYTORCHGEOMETRIC support was in development at that time and is now available.
With \TENSORFLOW custom backend available for the server, we can easily run DeepMET and DeepTau inference with the SONIC approach in \CMSSW on IPUs.

First, a per-algorithm throughput scan was carried out.
Comparing running inference on MK2 GC200 IPUs with NVIDIA Tesla V100s, a chip-to-chip factor of 3 throughput improvement was found, with larger gains expected for more computing-intensive models.
Running the entire Mini-AOD workflow was also tested.
We adapted the workflow configuration to point the \CMSSW clients running at the Purdue Tier-2 cluster to IPU servers on the Graphcloud cloud.
Without any other modifications on the client side, the workflow ran successfully, performing inference of the DeepMET and DeepTau models on the cloud and the other parts of the workflow on the client local CPUs.
Outputs were checked and found to be consistent with direct inference, within $10^{-6}$ differences for floats due to precision limits.

\section{Summary}
\label{sec:summary}
Within the next decade, the data-taking rate at the LHC will increase dramatically, straining the expected computing resources of the LHC experiments.
At the same time, more algorithms that run on these resources will be converted into either machine learning or domain algorithms that are easily accelerated with the use of coprocessors, such as graphics processing units (GPUs).
By pursuing heterogeneous architectures, it is possible to alleviate potential shortcomings of available central processing unit (CPU) resources.

Inference as a service (IaaS) is a promising scheme to integrate coprocessors into CMS computing workflows.
In IaaS, client code simply assembles the input data for an algorithm, sends that input to an inference server running either locally or remotely, and retrieves output from the server.
The implementation of IaaS discussed throughout this paper is called the Services for Optimized Network Inference on Coprocessors (SONIC) approach, which employs NVIDIA \TRITON Inference Servers to host models on coprocessors, as demonstrated here in studies on GPUs, CPUs, and Graphcore Intelligence Processing Units (IPUs).

In this paper, the SONIC approach in the CMS software framework (\CMSSW) is demonstrated in a sample Mini-AOD workflow, where algorithms for jet tagging, tau lepton identification, and missing transverse momentum regression are ported to run on inference servers.
These algorithms account for nearly 10\% of the total processing time per event in a simulated data set of top quark-antiquark events.
After model profiling, which is used to optimize server performance and determine the needed number of GPUs for a given number of client jobs, the expected 10\% decrease in per-event processing time was achieved in a large-scale test of Mini-AOD production with the SONIC approach that used about 10\,000 CPU cores and 100 GPUs.
The network bandwidth is large enough to support high input-output model inference for the workflow tested, and it will be monitored as the fraction of algorithms using remote GPUs increases.

In addition to meeting performance expectations, we demonstrated that the throughput results are not highly sensitive to the physical client-to-server distance, at least up to distances of hundreds of kilometers. Running inference through \TRITON servers on local CPU resources does not affect the throughput compared with the standard approach of running inference directly on CPUs in the job thread.
We also performed a test using GraphCore IPUs to demonstrate the flexibility of the SONIC approach.

The SONIC approach for IaaS represents a flexible method to accelerate algorithms, which is increasingly valuable for LHC experiments.
Using a realistic workflow, we highlighted many of the benefits of the SONIC approach, including the use of remote resources, workflow acceleration, and portability to different processor technologies.
To make it a viable and robust paradigm for CMS computing in the future, additional studies are ongoing or planned for monitoring and mitigating potential issues such as excessive network and memory usage or server failures.

\begin{acknowledgments}
    We congratulate our colleagues in the CERN accelerator departments for the excellent performance of the LHC and thank the technical and administrative staffs at CERN and at other CMS institutes for their contributions to the success of the CMS effort. In addition, we gratefully acknowledge the computing centres and personnel of the Worldwide LHC Computing Grid and other centres for delivering so effectively the computing infrastructure essential to our analyses. Finally, we acknowledge the enduring support for the construction and operation of the LHC, the CMS detector, and the supporting computing infrastructure provided by the following funding agencies: SC (Armenia), BMBWF and FWF (Austria); FNRS and FWO (Belgium); CNPq, CAPES, FAPERJ, FAPERGS, and FAPESP (Brazil); MES and BNSF (Bulgaria); CERN; CAS, MoST, and NSFC (China); MINCIENCIAS (Colombia); MSES and CSF (Croatia); RIF (Cyprus); SENESCYT (Ecuador); ERC PRG, RVTT3 and TK202 (Estonia); Academy of Finland, MEC, and HIP (Finland); CEA and CNRS/IN2P3 (France); SRNSF (Georgia); BMBF, DFG, and HGF (Germany); GSRI (Greece); NKFIH (Hungary); DAE and DST (India); IPM (Iran); SFI (Ireland); INFN (Italy); MSIP and NRF (Republic of Korea); MES (Latvia); LAS (Lithuania); MOE and UM (Malaysia); BUAP, CINVESTAV, CONACYT, LNS, SEP, and UASLP-FAI (Mexico); MOS (Montenegro); MBIE (New Zealand); PAEC (Pakistan); MES and NSC (Poland); FCT (Portugal); MESTD (Serbia); MCIN/AEI and PCTI (Spain); MOSTR (Sri Lanka); Swiss Funding Agencies (Switzerland); MST (Taipei); MHESI and NSTDA (Thailand); TUBITAK and TENMAK (Turkey); NASU (Ukraine); STFC (United Kingdom); DOE and NSF (USA).

    \hyphenation{Rachada-pisek} Individuals have received support from the Marie-Curie programme and the European Research Council and Horizon 2020 Grant, contract Nos.\ 675440, 724704, 752730, 758316, 765710, 824093, and COST Action CA16108 (European Union); the Leventis Foundation; the Alfred P.\ Sloan Foundation; the Alexander von Humboldt Foundation; the Science Committee, project no. 22rl-037 (Armenia); the Belgian Federal Science Policy Office; the Fonds pour la Formation \`a la Recherche dans l'Industrie et dans l'Agriculture (FRIA-Belgium); the Agentschap voor Innovatie door Wetenschap en Technologie (IWT-Belgium); the F.R.S.-FNRS and FWO (Belgium) under the ``Excellence of Science -- EOS" -- be.h project n.\ 30820817; the Beijing Municipal Science \& Technology Commission, No. Z191100007219010 and Fundamental Research Funds for the Central Universities (China); the Ministry of Education, Youth and Sports (MEYS) of the Czech Republic; the Shota Rustaveli National Science Foundation, grant FR-22-985 (Georgia); the Deutsche Forschungsgemeinschaft (DFG), under Germany's Excellence Strategy -- EXC 2121 ``Quantum Universe" -- 390833306, and under project number 400140256 - GRK2497; the Hellenic Foundation for Research and Innovation (HFRI), Project Number 2288 (Greece); the Hungarian Academy of Sciences, the New National Excellence Program - \'UNKP, the NKFIH research grants K 124845, K 124850, K 128713, K 128786, K 129058, K 131991, K 133046, K 138136, K 143460, K 143477, 2020-2.2.1-ED-2021-00181, and TKP2021-NKTA-64 (Hungary); the Council of Science and Industrial Research, India; ICSC -- National Research Centre for High Performance Computing, Big Data and Quantum Computing, funded by the EU NexGeneration program (Italy); the Latvian Council of Science; the Ministry of Education and Science, project no. 2022/WK/14, and the National Science Center, contracts Opus 2021/41/B/ST2/01369 and 2021/43/B/ST2/01552 (Poland); the Funda\c{c}\~ao para a Ci\^encia e a Tecnologia, grant CEECIND/01334/2018 (Portugal); the National Priorities Research Program by Qatar National Research Fund; MCIN/AEI/10.13039/501100011033, ERDF ``a way of making Europe", and the Programa Estatal de Fomento de la Investigaci{\'o}n Cient{\'i}fica y T{\'e}cnica de Excelencia Mar\'{\i}a de Maeztu, grant MDM-2017-0765 and Programa Severo Ochoa del Principado de Asturias (Spain); the Chulalongkorn Academic into Its 2nd Century Project Advancement Project, and the National Science, Research and Innovation Fund via the Program Management Unit for Human Resources \& Institutional Development, Research and Innovation, grant B37G660013 (Thailand); the Kavli Foundation; the Nvidia Corporation; the SuperMicro Corporation; the Welch Foundation, contract C-1845; and the Weston Havens Foundation (USA).
\end{acknowledgments}

\bibliography{auto_generated}
\cleardoublepage \appendix\section{The CMS Collaboration \label{app:collab}}\begin{sloppypar}\hyphenpenalty=5000\widowpenalty=500\clubpenalty=5000
\cmsinstitute{Yerevan Physics Institute, Yerevan, Armenia}
{\tolerance=6000
A.~Hayrapetyan, A.~Tumasyan\cmsAuthorMark{1}\cmsorcid{0009-0000-0684-6742}
\par}
\cmsinstitute{Institut f\"{u}r Hochenergiephysik, Vienna, Austria}
{\tolerance=6000
W.~Adam\cmsorcid{0000-0001-9099-4341}, J.W.~Andrejkovic, T.~Bergauer\cmsorcid{0000-0002-5786-0293}, S.~Chatterjee\cmsorcid{0000-0003-2660-0349}, K.~Damanakis\cmsorcid{0000-0001-5389-2872}, M.~Dragicevic\cmsorcid{0000-0003-1967-6783}, P.S.~Hussain\cmsorcid{0000-0002-4825-5278}, M.~Jeitler\cmsAuthorMark{2}\cmsorcid{0000-0002-5141-9560}, N.~Krammer\cmsorcid{0000-0002-0548-0985}, A.~Li\cmsorcid{0000-0002-4547-116X}, D.~Liko\cmsorcid{0000-0002-3380-473X}, I.~Mikulec\cmsorcid{0000-0003-0385-2746}, J.~Schieck\cmsAuthorMark{2}\cmsorcid{0000-0002-1058-8093}, R.~Sch\"{o}fbeck\cmsorcid{0000-0002-2332-8784}, D.~Schwarz\cmsorcid{0000-0002-3821-7331}, M.~Sonawane\cmsorcid{0000-0003-0510-7010}, S.~Templ\cmsorcid{0000-0003-3137-5692}, W.~Waltenberger\cmsorcid{0000-0002-6215-7228}, C.-E.~Wulz\cmsAuthorMark{2}\cmsorcid{0000-0001-9226-5812}
% ...
\par}
\vskip\cmsinstskip
\dag:~Deceased\\
$^{1}$Also at Yerevan State University, Yerevan, Armenia\\
$^{2}$Also at TU Wien, Vienna, Austria\\
% ...
\end{sloppypar}
\end{document}